\documentclass[10pt,journal,comsoc, tikz]{IEEEtran}

\usepackage[pdftex]{graphicx}
\usepackage{multirow}   
\usepackage{amsmath}
\usepackage{amssymb}
\usepackage{bbm}
\usepackage{eqparbox}
\usepackage{subcaption}
\usepackage{booktabs}
\usepackage{xtab}
\usepackage[hidelinks]{hyperref}
\usepackage{mathtools}
\usepackage{xcolor}
\usepackage{comment}
\usepackage{makecell}
\usepackage[acronyms,nonumberlist,nopostdot,nomain,nogroupskip,acronymlists={hidden}]{glossaries}
\usepackage[linesnumbered,ruled,vlined]{algorithm2e}
\usepackage{float} 
\usepackage[utf8]{inputenc}

\DeclareMathOperator{\diag}{diag}

\newacronym{3gpp}{3GPP}{3rd Generation Partnership Project}
\newacronym{4g}{4G}{4th generation}
\newacronym{5g}{5G}{5th generation}
\newacronym{6g}{6G}{6th generation}
\newacronym{5gc}{5GC}{5G Core}
\newacronym{adc}{ADC}{Analog to Digital Converter}
\newacronym{aerpaw}{AERPAW}{Aerial Experimentation and Research Platform for Advanced Wireless}
\newacronym{ai}{AI}{Artificial Intelligence}
\newacronym{LE}{LE}{Log-Euclidean}
\newacronym{TP}{TP}{Test Point}
\newacronym{CMD}{CMD}{Colinearity Matrix Distance}

\newacronym{EM}{EM}{electromagnetic}
\newacronym{EMS}{EMS}{electromagnetic skin}
\newacronym{mmW}{mmW}{millimeter wave}
\newacronym{RoI}{RoI}{region of interest}
\newacronym{AoA}{AoA}{Angle of Arrival}
\newacronym{AoD}{AoD}{Angle of Departure}
\newacronym{AoR}{AoR}{Angle of Reflection}
\newacronym{AoI}{AoI}{Angle of Incidence}
\newacronym{RIS}{RIS}{Reconfigurable Intelligent Surface}
\newacronym{RISs}{RISs}{Reconfigurable Intelligent Surfaces}
\newacronym{SNR}{SNR}{signal-to-noise ratio}
\newacronym{AF}{AF}{amplify-and-forward}
\newacronym{DF}{DF}{decode-and-forward}
\newacronym{STAR-RISs}{STAR-RISs}{Simultaneous Transmit and Reflecting RISs}
\newacronym{3GPP}{3GPP}{3rd Generation Partnership Project}
\newacronym{RAN}{RAN}{Radio Access Network}
\newacronym{KS}{KS}{Kruskal Stress}
\newacronym{CT}{CT}{Continuity}
\newacronym{TW}{TW}{Trustworthiness }
\newacronym{NCR}{NCR}{Network-Controlled Repeater}
\newacronym{NCRs}{NCRs}{Network-Controlled Repeaters}
\newacronym{IAB}{IAB}{Integated-Access-and-Backhauling}
\newacronym{UAV}{UAV}{Unmanned Aerial Vehicle}
\newacronym{SRE}{SRE}{Smart Radio Environment}
\newacronym{ECDF}{ECDF}{Empirical Cumulative Distribution Function}
\newacronym{CDF}{CDF}{Cumulative Distribution Function}
\newacronym{HSRE}{HSRE}{Heterogeneous SRE}
\newacronym{aimd}{AIMD}{Additive Increase Multiplicative Decrease}
\newacronym{am}{AM}{Acknowledged Mode}
\newacronym{GSTC}{GSTC}{Generalized Sheet Transition Condition}
\newacronym{amc}{AMC}{Adaptive Modulation and Coding}
\newacronym{FCAE}{FCAE}{Fully Connected Autoencoder}
\newacronym{amf}{AMF}{Access and Mobility Management Function}
\newacronym{aops}{AOPS}{Adaptive Order Prediction Scheduling}
\newacronym{api}{API}{Application Programming Interface}
\newacronym{apn}{APN}{Access Point Name}
\newacronym{LoS}{LoS}{Line-of-Sight}
\newacronym{NLoS}{NLoS}{None-Line-of-Sight}
\newacronym{NLDR}{NLDR}{Nonlinear Dimensionality Reduction}
\newacronym{NNs}{NNs}{Neural networks}

\newacronym{ap}{AP}{Application Protocol}
\newacronym{ae}{AE}{Autoencoder}
\newacronym{aqm}{AQM}{Active Queue Management}
\newacronym{ar}{AR}{Augmented Reality}
\newacronym{6G}{6G}{sixth-generation}
\newacronym{MLE}{MLE}{Mean Localization Error}
\newacronym{ausf}{AUSF}{Authentication Server Function}
\newacronym{avc}{AVC}{Advanced Video Coding}
\newacronym{awgn}{AGWN}{Additive White Gaussian Noise}
\newacronym{balia}{BALIA}{Balanced Link Adaptation Algorithm}
\newacronym{bbu}{BBU}{Base Band Unit}
\newacronym{bdp}{BDP}{Bandwidth-Delay Product}
\newacronym{ber}{BER}{Bit Error Rate}
\newacronym{bf}{BF}{Beamforming}
\newacronym{bler}{BLER}{Block Error Rate}
\newacronym{OSM}{OSM}{Open Street Map}
\newacronym{MSE}{MSE}{Mean Square Error}
\newacronym{brr}{BRR}{Bayesian Ridge Regressor}
\newacronym{BS}{BS}{Base Station}
\newacronym{bsr}{BSR}{Buffer Status Report}
\newacronym{bss}{BSS}{Business Support System}
\newacronym{ca}{CA}{Carrier Aggregation}
\newacronym{caas}{CaaS}{Connectivity-as-a-Service}
\newacronym{GPS}{GPS}{Global Positioning System}
\newacronym{cb}{CB}{Code Block}
\newacronym{cc}{CC}{channel charting}
\newacronym{ccid}{CCID}{Congestion Control ID}
\newacronym{cco}{CC}{Carrier Component}
\newacronym{cdd}{CDD}{Cyclic Delay Diversity}

\newacronym{cdn}{CDN}{Content Distribution Network}
\newacronym{cn}{CN}{Core Network}
\newacronym{codel}{CoDel}{Controlled Delay Management}
\newacronym{comac}{COMAC}{Converged Multi-Access and Core}
\newacronym{cord}{CORD}{Central Office Re-architected as a Datacenter}
\newacronym{cornet}{CORNET}{COgnitive Radio NETwork}
\newacronym{cosmos}{COSMOS}{Cloud Enhanced Open Software Defined Mobile Wireless Testbed for City-Scale Deployment}
\newacronym{cots}{COTS}{Commercial Off-the-Shelf}
\newacronym{cp}{CP}{Control Plane}
\newacronym{cyp}{CP}{Cyclic Prefix}
\newacronym{up}{UP}{User Plane}
\newacronym{cpu}{CPU}{Central Processing Unit}
\newacronym{cqi}{CQI}{Channel Quality Information}
\newacronym{cr}{CR}{Cognitive Radio}
\newacronym{cran}{C-RAN}{Cloud \gls{ran}}
\newacronym{crs}{CRS}{Cell Reference Signal}
\newacronym{csi}{CSI}{Channel State Information}
\newacronym{csirs}{CSI-RS}{Channel State Information - Reference Signal}
\newacronym{cu}{CU}{Central Unit}
\newacronym{d2tcp}{D$^2$TCP}{Deadline-aware Data center TCP}
\newacronym{d3}{D$^3$}{Deadline-Driven Delivery}
\newacronym{dac}{DAC}{Digital to Analog Converter}
\newacronym{dag}{DAG}{Directed Acyclic Graph}
\newacronym{das}{DAS}{Distributed Antenna System}
\newacronym{dash}{DASH}{Dynamic Adaptive Streaming over HTTP}
\newacronym{dc}{DC}{Dual Connectivity}
\newacronym{dccp}{DCCP}{Datagram Congestion Control Protocol}
\newacronym{dce}{DCE}{Direct Code Execution}
\newacronym{dci}{DCI}{Downlink Control Information}
\newacronym{dctcp}{DCTCP}{Data Center TCP}
\newacronym{dl}{DL}{Downlink}
\newacronym{dmr}{DMR}{Deadline Miss Ratio}
\newacronym{dmrs}{DMRS}{DeModulation Reference Signal}
\newacronym{drlcc}{DRL-CC}{Deep Reinforcement Learning Congestion Control}
\newacronym{drs}{DRS}{Discovery Reference Signal}
\newacronym{du}{DU}{Distributed Unit}
\newacronym{e2e}{E2E}{end-to-end}
\newacronym{ecaas}{ECaaS}{Edge-Cloud-as-a-Service}
\newacronym{ecn}{ECN}{Explicit Congestion Notification}
\newacronym{edf}{EDF}{Earliest Deadline First}
\newacronym{embb}{eMBB}{Enhanced Mobile Broadband}
\newacronym{empower}{EMPOWER}{EMpowering transatlantic PlatfOrms for advanced WirEless Research}
\newacronym{enb}{eNB}{evolved Node Base}
\newacronym{endc}{EN-DC}{E-UTRAN-\gls{nr} \gls{dc}}
\newacronym{epc}{EPC}{Evolved Packet Core}
\newacronym{eps}{EPS}{Evolved Packet System}
\newacronym{es}{ES}{Edge Server}
\newacronym{etsi}{ETSI}{European Telecommunications Standards Institute}
\newacronym[firstplural=Estimated Times of Arrival (ETAs)]{eta}{ETA}{Estimated Time of Arrival}
\newacronym{eutran}{E-UTRAN}{Evolved Universal Terrestrial Access Network}
\newacronym{faas}{FaaS}{Function-as-a-Service}
\newacronym{fapi}{FAPI}{Functional Application Platform Interface}
\newacronym{fdd}{FDD}{Frequency Division Duplexing}
\newacronym{fdm}{FDM}{Frequency Division Multiplexing}
\newacronym{fdma}{FDMA}{Frequency Division Multiple Access}
\newacronym{fed4fire}{FED4FIRE+}{Federation 4 Future Internet Research and Experimentation Plus}
\newacronym{fir}{FIR}{Finite Impulse Response}
\newacronym{fit}{FIT}{Future \acrlong{iot}}
\newacronym{fpga}{FPGA}{Field Programmable Gate Array}
\newacronym{fr2}{FR2}{Frequency Range 2}
\newacronym{fs}{FS}{Fast Switching}
\newacronym{fscc}{FSCC}{Flow Sharing Congestion Control}
\newacronym{ftp}{FTP}{File Transfer Protocol}
\newacronym{fw}{FW}{Flow Window}
\newacronym{ge}{GE}{Gaussian Elimination}
\newacronym{gnb}{gNB}{Next Generation Node Base}
\newacronym{gop}{GOP}{Group of Pictures}
\newacronym{gpr}{GPR}{Gaussian Process Regressor}
\newacronym{gpu}{GPU}{Graphics Processing Unit}
\newacronym{gtp}{GTP}{GPRS Tunneling Protocol}
\newacronym{gtpc}{GTP-C}{GPRS Tunnelling Protocol Control Plane}
\newacronym{gtpu}{GTP-U}{GPRS Tunnelling Protocol User Plane}
\newacronym{gtpv2c}{GTPv2-C}{\gls{gtp} v2 - Control}
\newacronym{gw}{GW}{Gateway}
\newacronym{harq}{HARQ}{Hybrid Automatic Repeat reQuest}
\newacronym{hetnet}{HetNet}{Heterogeneous Network}
\newacronym{hh}{HH}{Hard Handover}
\newacronym{hol}{HOL}{Head-of-Line}
\newacronym{hqf}{HQF}{Highest-quality-first}
\newacronym{hss}{HSS}{Home Subscription Server}
\newacronym{http}{HTTP}{HyperText Transfer Protocol}
\newacronym{ia}{IA}{Initial Access}
\newacronym{iab}{IAB}{Integrated Access and Backhaul}
\newacronym{ic}{IC}{Incident Command}
\newacronym{ietf}{IETF}{Internet Engineering Task Force}
\newacronym{imsi}{IMSI}{International Mobile Subscriber Identity}
\newacronym{imt}{IMT}{International Mobile Telecommunication}
\newacronym{iot}{IoT}{Internet of Things}
\newacronym{ip}{IP}{Internet Protocol}
\newacronym{itu}{ITU}{International Telecommunication Union}
\newacronym{kpi}{KPI}{Key Performance Indicator}
\newacronym{kpm}{KPM}{Key Performance Measurement}
\newacronym{kvm}{KVM}{Kernel-based Virtual Machine}
\newacronym{los}{LOS}{Line-of-Sight}
\newacronym{lsm}{LSM}{Link-to-System Mapping}
\newacronym{lstm}{LSTM}{Long Short Term Memory}
\newacronym{lte}{LTE}{Long Term Evolution}
\newacronym{lxc}{LXC}{Linux Container}
\newacronym{m2m}{M2M}{Machine to Machine}
\newacronym{mac}{MAC}{Medium Access Control}
\newacronym{manet}{MANET}{Mobile Ad Hoc Network}
\newacronym{mano}{MANO}{Management and Orchestration}
\newacronym{mc}{MC}{Multi-Connectivity}
\newacronym{mcc}{MCC}{Mobile Cloud Computing}
\newacronym{mchem}{MCHEM}{Massive Channel Emulator}
\newacronym{mcs}{MCS}{Modulation and Coding Scheme}
\newacronym{mec}{MEC}{Multi-access Edge Computing}
\newacronym{mec2}{MEC}{Mobile Edge Cloud}
\newacronym{mfc}{MFC}{Mobile Fog Computing}
\newacronym{mgen}{MGEN}{Multi-Generator}
\newacronym{mi}{MI}{Mutual Information}
\newacronym{mib}{MIB}{Master Information Block}
\newacronym{miesm}{MIESM}{Mutual Information Based Effective SINR}
\newacronym{mimo}{MIMO}{Multiple Input, Multiple Output}
\newacronym{ml}{ML}{Machine Learning}
\newacronym{mlr}{MLR}{Maximum-local-rate}
\newacronym[plural=\gls{mme}s,firstplural=Mobility Management Entities (MMEs)]{mme}{MME}{Mobility Management Entity}
\newacronym{mmtc}{mMTC}{Massive Machine-Type Communications}
\newacronym{mmwave}{mmWave}{millimeter wave}
\newacronym{mpdccp}{MP-DCCP}{Multipath Datagram Congestion Control Protocol}
\newacronym{mptcp}{MPTCP}{Multipath TCP}
\newacronym{mr}{MR}{Maximum Rate}
\newacronym{mrdc}{MR-DC}{Multi \gls{rat} \gls{dc}}
\newacronym{mse}{MSE}{Mean Square Error}
\newacronym{mss}{MSS}{Maximum Segment Size}
\newacronym{mt}{MT}{Mobile Termination}
\newacronym{mtd}{MTD}{Machine-Type Device}
\newacronym{mtu}{MTU}{Maximum Transmission Unit}
\newacronym{mumimo}{MU-MIMO}{Multi-user \gls{mimo}}
\newacronym{mvno}{MVNO}{Mobile Virtual Network Operator}
\newacronym{nalu}{NALU}{Network Abstraction Layer Unit}
\newacronym{nas}{NAS}{Non-Access Stratum}
\newacronym{nbiot}{NB-IoT}{Narrow Band IoT}
\newacronym{nfv}{NFV}{Network Function Virtualization}
\newacronym{nfvi}{NFVI}{Network Function Virtualization Infrastructure}
\newacronym{ngrg}{nGRG}{next Generation Research Group}
\newacronym{ni}{NI}{Network Interfaces}
\newacronym{nic}{NIC}{Network Interface Card}
\newacronym{nlos}{NLOS}{Non-Line-of-Sight}
\newacronym{now}{NOW}{Non Overlapping Window}
\newacronym{nsm}{NSM}{Network Service Mesh}
\newacronym{nr}{NR}{New Radio}
\newacronym{nrf}{NRF}{Network Repository Function}
\newacronym{nsa}{NSA}{Non Stand Alone}
\newacronym{nse}{NSE}{Network Slicing Engine}
\newacronym{nssf}{NSSF}{Network Slice Selection Function}
\newacronym{o2i}{O2I}{Outdoor to Indoor}
\newacronym{oai}{OAI}{OpenAirInterface}
\newacronym{oaicn}{OAI-CN}{\gls{oai} \acrlong{cn}}
\newacronym{oairan}{OAI-RAN}{\acrlong{oai} \acrlong{ran}}
\newacronym{oam}{OAM}{Operations, Administration and Maintenance}
\newacronym{ofdm}{OFDM}{Orthogonal Frequency Division Multiplexing}
\newacronym{olia}{OLIA}{Opportunistic Linked Increase Algorithm}
\newacronym{omec}{OMEC}{Open Mobile Evolved Core}
\newacronym{onap}{ONAP}{Open Network Automation Platform}
\newacronym{onf}{ONF}{Open Networking Foundation}
\newacronym{onos}{ONOS}{Open Networking Operating System}
\newacronym{oom}{OOM}{\gls{onap} Operations Manager}
\newacronym{opnfv}{OPNFV}{Open Platform for \gls{nfv}}
\newacronym{oran}{O-RAN}{Open Radio Access Network}
\newacronym{orbit}{ORBIT}{Open-Access Research Testbed for Next-Generation Wireless Networks}
\newacronym{os}{OS}{Operating System}
\newacronym{oss}{OSS}{Operations Support System}
\newacronym{otic}{OTIC}{Open Testing \& Integration Centre}
\newacronym{pa}{PA}{Position-aware}
\newacronym{pase}{PASE}{Prioritization, Arbitration, and Self-adjusting Endpoints}
\newacronym{pawr}{PAWR}{Platforms for Advanced Wireless Research}
\newacronym{pbch}{PBCH}{Physical Broadcast Channel}
\newacronym{pcef}{PCEF}{Policy and Charging Enforcement Function}
\newacronym{pcfich}{PCFICH}{Physical Control Format Indicator Channel}
\newacronym{pcrf}{PCRF}{Policy and Charging Rules Function}
\newacronym{pdcch}{PDCCH}{Physical Downlink Control Channel}
\newacronym{pdcp}{PDCP}{Packet Data Convergence Protocol}
\newacronym{pdsch}{PDSCH}{Physical Downlink Shared Channel}
\newacronym{pdu}{PDU}{Packet Data Unit}
\newacronym{pf}{PF}{Proportional Fair}
\newacronym{pgw}{PGW}{Packet Gateway}
\newacronym{phich}{PHICH}{Physical Hybrid ARQ Indicator Channel}
\newacronym{phy}{PHY}{Physical}
\newacronym{pmch}{PMCH}{Physical Multicast Channel}
\newacronym{pmi}{PMI}{Precoding Matrix Indicators}
\newacronym{powder}{POWDER}{Platform for Open Wireless Data-driven Experimental Research}
\newacronym{ppo}{PPO}{Proximal Policy Optimization}
\newacronym{ppp}{PPP}{Poisson Point Process}
\newacronym{prach}{PRACH}{Physical Random Access Channel}
\newacronym{prb}{PRB}{Physical Resource Block}
\newacronym{psnr}{PSNR}{Peak Signal to Noise Ratio}
\newacronym{pss}{PSS}{Primary Synchronization Signal}
\newacronym{pucch}{PUCCH}{Physical Uplink Control Channel}
\newacronym{pusch}{PUSCH}{Physical Uplink Shared Channel}
\newacronym{rar}{RAR}{Random Access Response}
\newacronym{qam}{QAM}{Quadrature Amplitude Modulation}
\newacronym{qci}{QCI}{\gls{qos} Class Identifier}
\newacronym{5qi}{5QI}{5G \gls{qos} Identifier}
\newacronym{qoe}{QoE}{Quality of Experience}
\newacronym{QoS}{QoS}{Quality of Service}
\newacronym{UE}{UE}{User Equipment}
\newacronym{UEs}{UEs}{User Equipments}
\newacronym{FoV}{FoV}{field of view}
\newacronym{UPA}{UPA}{uniform planar array}
\newacronym{quic}{QUIC}{Quick UDP Internet Connections}
\newacronym{rach}{RACH}{Random Access Channel}
\newacronym{ran}{RAN}{Radio Access Network}
\newacronym[firstplural=Radio Access Technologies (RATs)]{rat}{RAT}{Radio Access Technology}
\newacronym{rcn}{RCN}{Research Coordination Network}
\newacronym{STAR}{STAR-RIS}{simultaneous transmitting and reflecting RIS}
\newacronym{3SNCR}{3SNCR}{trisectoral NCR}
\newacronym{rc}{RC}{RAN Control}
\newacronym{rec}{REC}{Radio Edge Cloud}
\newacronym{red}{RED}{Random Early Detection}
\newacronym{renew}{RENEW}{Reconfigurable Eco-system for Next-generation End-to-end Wireless}
\newacronym{rf}{RF}{Radio Frequency}
\newacronym{rfc}{RFC}{Request for Comments}
\newacronym{rfr}{RFR}{Random Forest Regressor}
\newacronym{ric}{RIC}{\gls{ran} Intelligent Controller}
\newacronym{rlc}{RLC}{Radio Link Control}
\newacronym{rlf}{RLF}{Radio Link Failure}
\newacronym{rlnc}{RLNC}{Random Linear Network Coding}
\newacronym{rmr}{RMR}{RIC Message Router}
\newacronym{rmse}{RMSE}{Root Mean Squared Error}
\newacronym{rnis}{RNIS}{Radio Network Information Service}
\newacronym{rr}{RR}{Round Robin}
\newacronym{rrc}{RRC}{Radio Resource Control}
\newacronym{rrm}{RRM}{Radio Resource Management}
\newacronym{rru}{RRU}{Remote Radio Unit}
\newacronym{rs}{RS}{Remote Server}
\newacronym{rsrp}{RSRP}{Reference Signal Received Power}
\newacronym{rsrq}{RSRQ}{Reference Signal Received Quality}
\newacronym{rss}{RSS}{Received Signal Strength}
\newacronym{rssi}{RSSI}{Received Signal Strength Indicator}
\newacronym{rtt}{RTT}{Round Trip Time}
\newacronym{ru}{RU}{Radio Unit}
\newacronym{rw}{RW}{Receive Window}
\newacronym{rx}{RX}{Receiver}
\newacronym{s1ap}{S1AP}{S1 Application Protocol}
\newacronym{sa}{SA}{standalone}
\newacronym{sack}{SACK}{Selective Acknowledgment}
\newacronym{sap}{SAP}{Service Access Point}
\newacronym{sc2}{SC2}{Spectrum Collaboration Challenge}
\newacronym{scef}{SCEF}{Service Capability Exposure Function}
\newacronym{sch}{SCH}{Secondary Cell Handover}
\newacronym{scoot}{SCOOT}{Split Cycle Offset Optimization Technique}
\newacronym{sctp}{SCTP}{Stream Control Transmission Protocol}
\newacronym{sdap}{SDAP}{Service Data Adaptation Protocol}
\newacronym{sdk}{SDK}{Software Development Kit}
\newacronym{sdm}{SDM}{Space Division Multiplexing}
\newacronym{sdma}{SDMA}{Spatial Division Multiple Access}
\newacronym{sdn}{SDN}{Software-defined Networking}
\newacronym{sdr}{SDR}{Software-defined Radio}
\newacronym{seba}{SEBA}{SDN-Enabled Broadband Access}
\newacronym{sgsn}{SGSN}{Serving GPRS Support Node}
\newacronym{sgw}{SGW}{Service Gateway}
\newacronym{si}{SI}{Study Item}
\newacronym{sib}{SIB}{Secondary Information Block}
\newacronym{sinr}{SINR}{Signal to Interference plus Noise Ratio}
\newacronym{sip}{SIP}{Session Initiation Protocol}
\newacronym{siso}{SISO}{Single Input, Single Output}
\newacronym{sla}{SLA}{Service Level Agreement}
\newacronym{sm}{SM}{Service Model}
\newacronym{smf}{SMF}{Session Management Function}
\newacronym{smo}{SMO}{Service Management and Orchestration}
\newacronym{sms}{SMS}{Short Message Service}
\newacronym{smsgmsc}{SMS-GMSC}{\gls{sms}-Gateway}
\newacronym{snr}{SNR}{Signal-to-Noise-Ratio}
\newacronym{son}{SON}{Self-Organizing Network}
\newacronym{sptcp}{SPTCP}{Single Path TCP}
\newacronym{srb}{SRB}{Service Radio Bearer}
\newacronym{srn}{SRN}{Standard Radio Node}
\newacronym{srs}{SRS}{Sounding Reference Signal}
\newacronym{zc}{ZC}{Zadoff-Chu}
\newacronym{ta}{TA}{Timing Advance}
\newacronym{ss}{SS}{Synchronization Signal}
\newacronym{sss}{SSS}{Secondary Synchronization Signal}
\newacronym{st}{ST}{Spanning Tree}
\newacronym{svc}{SVC}{Scalable Video Coding}
\newacronym{tb}{TB}{Transport Block}
\newacronym{tcp}{TCP}{Transmission Control Protocol}
\newacronym{tdd}{TDD}{Time Division Duplexing}
\newacronym{tdm}{TDM}{Time Division Multiplexing}
\newacronym{tdma}{TDMA}{Time Division Multiple Access}
\newacronym{tfl}{TfL}{Transport for London}
\newacronym{tfrc}{TFRC}{TCP-Friendly Rate Control}
\newacronym{tft}{TFT}{Traffic Flow Template}
\newacronym{tgen}{TGEN}{Traffic Generator}
\newacronym{tip}{TIP}{Telecom Infra Project}
\newacronym{tm}{TM}{Transparent Mode}
\newacronym{to}{TO}{Telco Operator}
\newacronym{tr}{TR}{Technical Report}
\newacronym{trp}{TRP}{Transmitter Receiver Pair}
\newacronym{ts}{TS}{Technical Specification}
\newacronym{tti}{TTI}{Transmission Time Interval}
\newacronym{ttt}{TTT}{Time-to-Trigger}
\newacronym{tx}{TX}{Transmitter}
\newacronym{uas}{UAS}{Unmanned Aerial System}
\newacronym{uav}{UAV}{Unmanned Aerial Vehicle}
\newacronym{udm}{UDM}{Unified Data Management}
\newacronym{udp}{UDP}{User Datagram Protocol}
\newacronym{udr}{UDR}{Unified Data Repository}
\newacronym{ue}{UE}{User Equipment}
\newacronym{uhd}{UHD}{\gls{usrp} Hardware Driver}
\newacronym{ul}{UL}{Uplink}
\newacronym{um}{UM}{Unacknowledged Mode}
\newacronym{uml}{UML}{Unified Modeling Language}
\newacronym{upa}{UPA}{Uniform Planar Array}
\newacronym{upf}{UPF}{User Plane Function}
\newacronym{urllc}{URLLC}{Ultra Reliable and Low Latency Communications}
\newacronym{usa}{U.S.}{United States}
\newacronym{usim}{USIM}{Universal Subscriber Identity Module}
\newacronym{usrp}{USRP}{Universal Software Radio Peripheral}
\newacronym{utc}{UTC}{Urban Traffic Control}
\newacronym{vim}{VIM}{Virtualization Infrastructure Manager}
\newacronym{vm}{VM}{Virtual Machine}
\newacronym{vnf}{VNF}{Virtual Network Function}
\newacronym{volte}{VoLTE}{Voice over \gls{lte}}
\newacronym{voltha}{VOLTHA}{Virtual OLT HArdware Abstraction}
\newacronym{vr}{VR}{Virtual Reality}
\newacronym{vran}{vRAN}{Virtualized \gls{ran}}
\newacronym{vss}{VSS}{Video Streaming Server}
\newacronym{wbf}{WBF}{Wired Bias Function}
\newacronym{wf}{WF}{Waterfilling}
\newacronym{wg}{WG}{Working Group}
\newacronym{wlan}{WLAN}{Wireless Local Area Network}
\newacronym{osm}{OSM}{Open Source Management and Orchestration}
\newacronym{pnf}{PNF}{Physical Network Function}
\newacronym{drl}{DRL}{Deep Reinforcement Learning}
\newacronym{mtc}{MTC}{Machine-type Communications}
\newacronym{osc}{OSC}{O-RAN Software Community}
\newacronym{mns}{MnS}{Management Services}
\newacronym{ves}{VES}{\gls{vnf} Event Stream}
\newacronym{ei}{EI}{Enrichment Information}
\newacronym{fh}{FH}{Fronthaul}
\newacronym{fft}{FFT}{Fast Fourier Transform}
\newacronym{laa}{LAA}{Licensed-Assisted Access}
\newacronym{plfs}{PLFS}{Physical Layer Frequency Signals}
\newacronym{ptp}{PTP}{Precision Time Protocol}
\newacronym{asic}{ASIC}{Application-specific Integrated Circuit}
\newacronym{aal}{AAL}{Acceleration Abstraction Layer}
\newacronym{fec}{FEC}{Forward Error Correction}
\newacronym{sdl}{SDL}{Shared Data Layer}
\newacronym{nib}{NIB}{Network Information Base}
\newacronym{rnib}{R-NIB}{RAN \gls{nib}}
\newacronym{fcaps}{FCAPS}{Fault, Configuration, Accounting, Performance, Security}
\newacronym{ie}{IE}{Information Element}
\newacronym{fg}{FG}{Focus Group}
\newacronym{osfg}{OSFG}{Open Source Focus Group}
\newacronym{sdfg}{SDFG}{Standard Development Focus Group}
\newacronym{tifg}{TIFG}{Test \& Integration Focus Group}
\newacronym{sfg}{SFG}{Security Focus Group}
\newacronym{swg}{SWG}{Security Work Group}
\newacronym{e2sm}{E2SM}{E2 Service Model}
\newacronym{tsc}{TSC}{Technical Steering Committee}
\newacronym{sdo}{SDO}{Standard-Development Organization}
\newacronym{sql}{SQL}{Structured Query Language}
\newacronym{ssh}{SSH}{Secure Shell}
\newacronym{tls}{TLS}{Transport Layer Security}
\newacronym{netconf}{NETCONF}{Network Configuration Protocol}
\newacronym{dtls}{DTLS}{Datagram Transport Layer Security}
\newacronym{cmp}{CMP}{Certificate Management Protocol}
\newacronym{ccc}{CCC}{Cell Configuration and Control}
\newacronym{dsp}{DSP}{Digital Signal Processing}
\newacronym{opex}{OPEX}{Operational Expenses}
\newacronym{cbrs}{CBRS}{Citizen Broadband Radio Service}
\newacronym{ntn}{NTN}{Non-terrestrial Network}
\newacronym{gbr}{GBR}{Guaranteed Bitrate}
\newacronym{sps}{SPS}{Semi-Persistent Scheduling}
\newacronym{tbs}{TBS}{Transport Block Size}
\newacronym{gnss}{GNSS}{Global Navigation Satellite System}
\newacronym{tof}{ToF}{Time of Flight}
\newacronym{rtof}{RToF}{Return Time of Flight}
\newacronym{rsig}{RS}{Reference Signal}
\newacronym{nrtric}{near-RT RIC}{near-Real Time Ran Intelligent Controller}
\newacronym{nonrtric}{non-RT RIC}{non-Real Time Ran Intelligent Controller}
\newacronym{aoa}{AoA}{Angle of Arrival}
\newacronym{tdoa}{TDoA}{Time Difference of Arrival}
\newacronym{rtoa}{RToA}{Return Time of Arrival}
\newacronym{ris}{RIS}{Reconfigurable Intelligent Surface}
\newacronym{srd}{SRD}{Smart Radio Device}
\newacronym{gfbr}{GFBR}{Guaranteed Flow Bit Rate}
\newacronym{rg}{RG}{Resource Grid}
\newacronym{rb}{RB}{Resource Block}
\newacronym{re}{RE}{Resource Element}
\newacronym{rfra}{RF}{Radio Frame}
\newacronym{scs}{SCS}{Subcarrier Spacing}
\newacronym{ec}{EC}{Edge Computing}
\newacronym{af}{AF}{Amplify-and-Forward}
\newacronym{ncr}{NCR}{Network-Controlled Repeater}
\newacronym{tp}{TP}{Test Point}
\newacronym{cs}{CS}{Candidate Site}
\newacronym{src}{SRC}{Smart Radio Connection}
\newacronym{milp}{MILP}{Mixed Integer-Linear Programming}

\newacronym{FCMC}{FCMC}{full coverage minimum cost}

\newacronym{MBCC}{MBCC}{maximum budget-constrained coverage}

\newacronym{PDF}{PDF}{probability density function}
\newacronym{tsne}{t-SNE}{t-distributed stochastic neighbor 
embedding}
\newacronym{stsne}{St-SNE}{Semi-Supervised t-distributed stochastic neighbor 
embedding}

\begin{document}

\title{Towards Channel Charting Enhancement with Non-Reconfigurable Intelligent Surfaces}

\author{Mahdi~Maleki,~\IEEEmembership{Graduate Student Member,~IEEE,} Reza~Agahzadeh~Ayoubi,~\IEEEmembership{Member,~IEEE,}
        Marouan~Mizmizi,~\IEEEmembership{Member,~IEEE,}
        Umberto~Spagnolini,~\IEEEmembership{Senior~Member,~IEEE}
       \thanks{This work was partially supported by the European Union - Next Generation EU under the Italian National Recovery and Resilience Plan (NRRP), Mission 4, Component 2, Investment 1.3, CUP D43C22003080001, partnership on “Telecommunications of the Future” (PE00000001 - program “RESTART”)}
\thanks{The authors are with the Department of Electronics, Information and Bioengineering, Politecnico di Milano, 20133, Milano, Italy}}

\maketitle

\begin{abstract}
We investigate how fully-passive \glspl{EMS} can be engineered
to enhance \gls{cc} in dense urban environments. We employ two
complementary state-of-the-art \gls{cc} techniques—semi-supervised
\gls{tsne} and a semi-supervised \gls{ae}—to verify the consistency
of results across nonparametric and parametric mappings.
We show that the accuracy of \gls{cc} hinges on a balance between
signal-to-noise ratio (SNR) and spatial dissimilarity:
\gls{EMS} codebooks that only maximize gain, as in conventional
\gls{RIS} optimization, suppress location fingerprints and
degrade \gls{cc}, while randomized phases increase diversity but
reduce SNR. To address this trade-off, we design static
\gls{EMS} phase profiles via a quantile-driven criterion that
targets worst-case users and improves both trustworthiness and
continuity. In a 3D ray-traced city at 30\,GHz, the proposed
\gls{EMS} reduces the 90th-percentile localization error from
$>$50\,m to $<$25\,m for both \gls{tsne}- and \gls{ae}-based
\gls{cc}, and decreases severe trajectory dropouts by over 4×
under 15\% supervision. The improvements hold consistently
across the evaluated configurations, establishing static,
pre-configured \glspl{EMS} as a practical enabler of \gls{cc}
without reconfiguration overheads.
\end{abstract}

\glsresetall
\begin{keywords}
Channel charting, electromagnetic skins, dissimilarity, Dimensionality reduction.
\end{keywords}

\glsresetall

\section{Introduction}

Wireless \gls{cc} has emerged as a powerful paradigm for exploiting the intrinsic characteristics of wireless propagation environments \cite{ferrand2023wireless}. By constructing a low-dimensional chart from high-dimensional \gls{csi}, \gls{cc} enables an unsupervised interpretation of the complex spatial and temporal structure of wireless channels. This data-driven framework supports a wide range of applications, including device localization, mobility tracking, and network optimization, and is increasingly regarded as a key enabler for future 6G networks \cite{mmWLocalization}.

Conventional localization techniques, such as \gls{GPS} or radio-access-based localization, have some limitations. Multipath propagation and \gls{nlos} conditions degrade the accuracy of these methods by causing signal distortions and delays, particularly in dense urban environments \cite{gupta2015survey}. Additionally, achieving precise localization often requires accurate synchronization between devices, finely calibrated hardware, and dedicated signaling (e.g., positioning reference signals \cite{ismail2019rssi}), increasing complexity and cost \cite{mmWLocalization}.

In contrast, \gls{cc} addresses these challenges by taking advantage of multipath propagation rather than being hindered by it. It extracts and organizes the rich spatial information inherently contained in \gls{csi}, allowing localization without the need for precise synchronization or specialized hardware calibration \cite{ferrand2023wireless}. Another practical advantage is that \gls{cc} can operate using existing communication infrastructures without requiring protocol modifications or additional signaling. It simply repurposes the \gls{csi} data already available in contemporary wireless systems \cite{taner2023channel}.

To obtain low-dimensional representations of \gls{csi}, \gls{cc} typically employs non-linear dimensionality reduction techniques that preserve the intrinsic geometric structure of the data. Popular approaches include Isomap~\cite{ISOmap} and \gls{tsne}~\cite{kazemi2023beam}. Among these, \gls{tsne} has gained considerable attention in \gls{cc} research due to its ability to reveal fine spatial relationships while maintaining a favorable balance between interpretability and performance. In parallel, \gls{ae}-based neural networks~\cite{studer2018channel,zhang2021semi,huang2019improving} have emerged as a powerful alternative, capable of learning complex non-linear mappings through data-driven training. Both \gls{tsne} and \gls{ae}-based models therefore provide effective means to project high-dimensional channel features into compact latent spaces that preserve the spatial relationships among users.

While these techniques effectively embed high-dimensional \gls{csi} into a low-dimensional latent space that reflects the spatial relationships among users, a critical challenge remains: to achieve device localization, one must map points in the latent space to actual physical coordinates. This mapping can be realized through supervised learning, where both \gls{csi} and corresponding user positions are available during training. However, collecting accurate position labels for all devices is generally impractical in real deployments due to cost, privacy, and logistical constraints~\cite{taner2025channel}.

To overcome this limitation, semi-supervised learning approaches~\cite{zhang2021semi} have been developed. In this paradigm, only a subset of devices—potentially dedicated training \gls{ue}s equipped with accurate positioning capabilities (e.g., \gls{GPS})—report their physical locations during the learning phase. The learned mapping is then generalized to infer the positions of all other devices from their \gls{csi} alone, significantly reducing the burden of position labeling while retaining localization accuracy.

High frequencies such as \gls{mmW} and upper mid-bands (6--24~GHz), are among the most promising candidates for 6G networks, as they provide the wide bandwidths required to support high-throughput and low-latency applications \cite{deng2021semi,kazemi2023beam,CIRS}. Nevertheless, applying \gls{cc} at these frequencies presents nontrivial challenges. The high attenuation and blockage typical of these bands lead to sparse and low-rank propagation channels \cite{mmWLocalization,shahmansoori2017position}, which in turn limit spatial diversity and hinder the ability of \gls{cc} to reliably distinguish between users located in close proximity \cite{studer2018channel}.

A practical strategy to alleviate these propagation limitations is offered by the concept of a \gls{SRE}, where the wireless environment is deliberately engineered to improve communication performance \cite{CEMS_V2V}. Within this framework, metasurfaces play a central role, as they are artificially engineered materials capable of controlling and manipulating electromagnetic waves \cite{MTM}. Among the different types of metasurfaces, the most widely studied are \glspl{RIS}. \glspl{RIS} can dynamically adjust their reflection properties to steer or reshape incident signals, effectively altering the characteristics of the wireless channel. Over the past decade, they have been extensively explored for several purposes, such as mitigating blockage \cite{Blockage1,RIS_vs_NCR}, enhancing coverage \cite{RIS_vs_NCR}, and enabling advanced functionalities \cite{Bennis_CC_for_RIS,RIS_RS_Eugenio}.

However, while \glspl{RIS} provide a high degree of control, their continuous reconfiguration and associated hardware infrastructure introduce significant cost and complexity \cite{ayanoglu2022wave}. More importantly, their operation requires prior knowledge of user locations to optimize reflections, creating a circular dependency that fundamentally conflicts with the goal of unsupervised localization. This limitation makes reconfigurable surfaces unsuitable for direct use within a \gls{cc} framework.

In contrast, fully passive \glspl{EMS} (namely smart skins) offer a simple and cost-effective alternative. These static, fully passive metasurfaces reflect incident waves according to the generalized Snell’s law \cite{SmartSkin}. Once deployed, they operate without any need for active control or feedback, providing a low-cost means to enrich the multipath structure of the environment. With an approximate cost of only a few dollars per square meter \cite{smsk_electronics}, \glspl{EMS} represent a practical solution for enhancing channel diversity and improving localization accuracy within the broader context of \gls{SRE}.

This work, as an extension of our previous study \cite{Maleki_CC_in_SRE}, advances \gls{cc} in smart radio environments by leveraging fully passive, static \glspl{EMS} to enrich multipath and improve the geometric fidelity of \gls{csi}-based embeddings. Our prior work introduced quantile-based optimization of static \gls{EMS} codebooks for \gls{cc} using \gls{tsne}. In this extended study, we broaden the scope by integrating a semi-supervised autoencoder to validate and complement the \gls{tsne} results across parametric and nonparametric mappings, and by conducting a more extensive and realistic evaluation that includes trajectory-based localization and a comparative analysis with an idealized \gls{RIS} and random static phases. Furthermore, we provide new numerical evidence that localization accuracy improves not through higher \gls{SNR} or stronger dissimilarity alone, but through a balanced trade-off between the two—an insight unique to this work.

It is worth noting that a recent study~\cite{CC_UAV} investigated \gls{cc} in a \gls{RIS}-assisted \gls{UAV} navigation context, where the \gls{RIS} configuration and channel state are fully known to the controller and used in a supervised manner. The goal there is not to improve the intrinsic accuracy of \gls{cc}, but to employ it as an auxiliary localization mechanism within a coordinated and labeled network. In contrast, our setting focuses on unsupervised and semi-supervised learning, where passive \glspl{EMS}—without reconfiguration or position feedback—enhance the geometric consistency of \gls{cc}. Similarly, other works such as~\cite{Bennis_CC_for_RIS} employ \gls{cc} to assist \gls{RIS} configuration, rather than the reverse—i.e., using metasurfaces to enhance \gls{cc}. 

The proposed framework operates in two stages: a learning phase, where \gls{csi} is collected for a subset of labeled test points, and an inference phase, where unseen user positions are estimated using the learned low-dimensional manifold.  

The main contributions of this work are summarized as follows:
\begin{itemize}
    \item We propose a unified framework for \gls{cc}-based localization in smart radio environments, where static \glspl{EMS} are optimized to enhance multipath diversity.
    \item We incorporate a semi-supervised \gls{ae}-based channel charting model to validate and generalize the results obtained with \gls{tsne}.
    \item We conduct a comprehensive analysis across different supervision levels, trajectories, and surface configurations, including comparisons with idealized \gls{RIS} and random static phases.
    \item We demonstrate numerically that localization accuracy improves through a balance between \gls{SNR} and channel dissimilarity, not by maximizing either individually.
\end{itemize}

Simulation results show that, in a representative city scenario, optimized \gls{EMS} configurations reduce the 90th-percentile localization error from over 50 meters (without \gls{EMS}) to less than 25 meters, while substantially improving \gls{TW} and \gls{CT} for the most challenging user locations. These gains are robust across both algorithmic choices and levels of supervision, and highlight the practical value of physically optimized \gls{EMS} in channel charting applications.

The remainder of the paper is organized as follows: Section II presents the system model and signal concepts. Section III reviews various \gls{cc} methods, while Section IV examines the impact of \gls{SRE} configuration on \gls{cc} performance and localization errors. Section V presents numerical results and discussion. Finally, conclusions are drawn from the findings.

\section{System Model}
\label{sec:system_model}

\begin{figure}[t!]
    \centering
    \includegraphics[width=\linewidth]{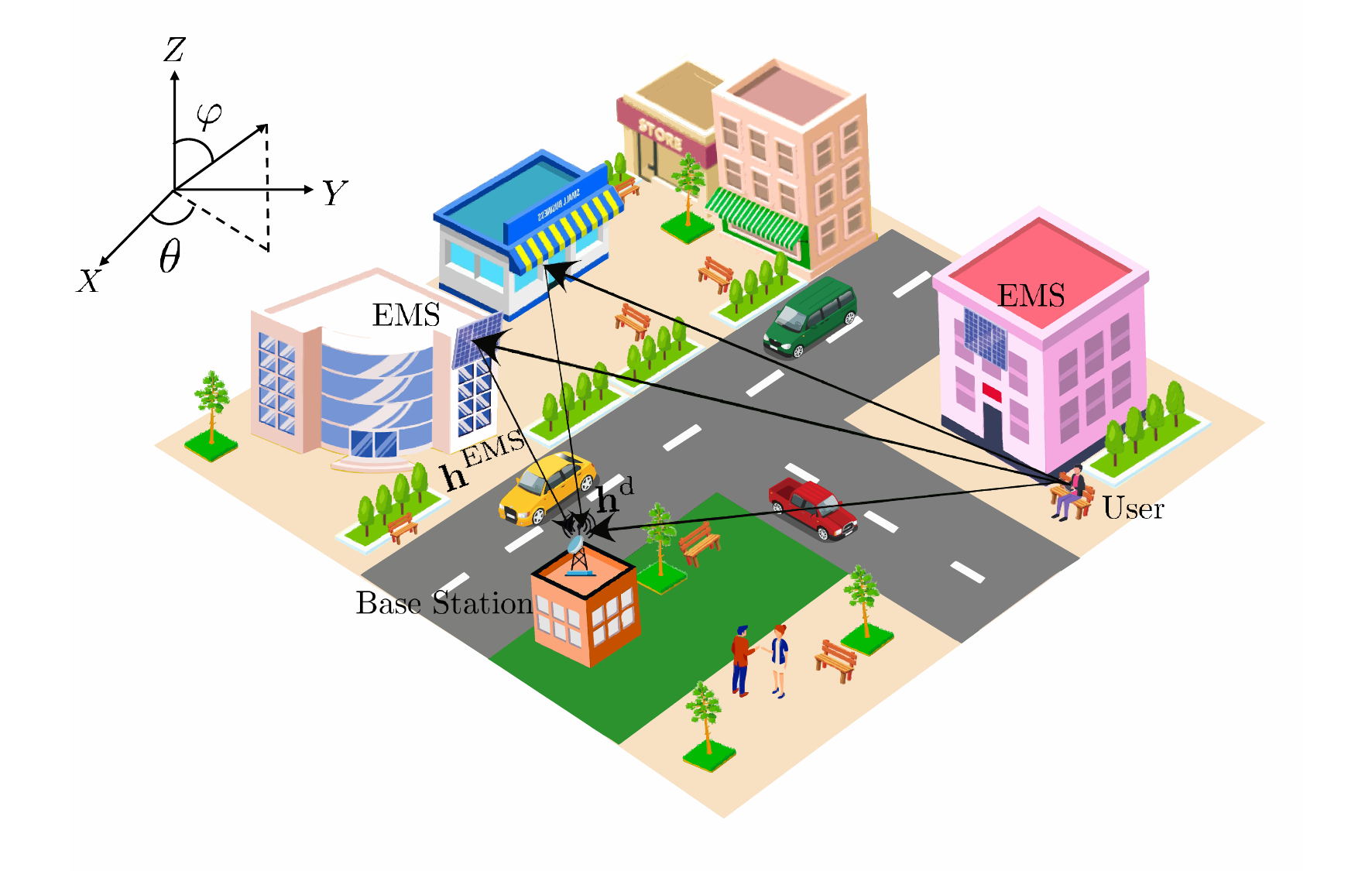}
    \caption{Reference scenario and system, adapted from \cite{Maleki_CC_in_SRE}}
    \label{fig:system model}
\end{figure}

Consider the wireless communication scenario illustrated in Fig.~\ref{fig:system model}, which comprises a \gls{BS} located at known position $\mathbf{P}_{\mathrm{BS}} \in \mathbb{R}^3$ and equipped with $N_{\mathrm{BS}}$ antennas, a single-antenna \gls{ue} at position $\mathbf{P}_{\mathrm{UE}} \in \mathbb{R}^3$, and a set $\mathcal{E}$ of $M = |\mathcal{E}|$ \glspl{EMS} (electromagnetic surfaces) placed at positions $\{\mathbf{P}_j^{\mathrm{EMS}}\}_{j=1}^M$, all expressed in a global reference system.
Each \gls{EMS} consists of $L$ sub-wavelength meta-atoms located at $\{\mathbf{p}_{j,\ell}^{\mathrm{EMS}}\}_{\ell=1}^{L}$ relative to the center $\mathbf{P}_j^{\mathrm{EMS}}$. 

The detailed procedure for channel charting and position inference from \gls{csi} will be described in Sec.~\ref{sec:cc_method}.

\subsection{Signal Model}
\label{subsec:signal_model}

Let $s \in \mathbb{C}$ be the transmit symbol from the \gls{ue} with $\mathbb{E}[|s|^2] = \sigma_s^2$. The time-discrete received signal at the \gls{BS} is modeled as
\begin{equation}
    \mathbf{y}_{\mathrm{rx}} = \mathbf{h}(\mathcal{S}) \, s + \mathbf{n},
    \label{eq:rx_signal}
\end{equation}
where $\mathbf{y}_{\mathrm{rx}} \in \mathbb{C}^{N_{\mathrm{BS}} \times 1}$ is the received vector, $\mathbf{n} \sim \mathcal{CN}(\mathbf{0}, \sigma_n^2 \mathbf{I}_{N_{\mathrm{BS}}})$ is the complex Gaussian noise, and $\mathbf{h}(\mathcal{S}) \in \mathbb{C}^{N_{\mathrm{BS}} \times 1}$ is the composite channel vector, which depends on the set of \gls{EMS} reflection configurations $\mathcal{S} = \{\boldsymbol{\Phi}_1, \dots, \boldsymbol{\Phi}_M\}$.

The channel vector is the superposition of the direct path and the reflected paths via all \gls{EMS}:
\begin{equation}
    \mathbf{h}(\mathcal{S}) = \mathbf{h}^{\mathrm{d}} + \sum_{j \in \mathcal{E}} \mathbf{h}_j^{\mathrm{EMS}}(\boldsymbol{\Phi}_j),
    \label{eq:total_channel}
\end{equation}
where $\mathbf{h}^{\mathrm{d}} \in \mathbb{C}^{N_{\mathrm{BS}} \times 1}$ is the direct channel between \gls{ue} and \gls{BS}, and $\mathbf{h}_j^{\mathrm{EMS}}(\boldsymbol{\Phi}_j) \in \mathbb{C}^{N_{\mathrm{BS}} \times 1}$ is the contribution of the $j$-th \gls{EMS}.
For simplicity, each EMS is configured by a diagonal reflection matrix $\boldsymbol{\Phi}_j \in \mathbb{C}^{L \times L}$:
\begin{equation}
    \boldsymbol{\Phi}_j = \mathrm{diag}\left( e^{j\phi_{j,1}}, \ldots, e^{j\phi_{j,L}} \right),
    \label{eq:ems_phi}
\end{equation}
where $\phi_{j,\ell}$ is the phase shift introduced by the $\ell$-th element of the $j$-th EMS. This modeling is widely adopted in the literature (see~\cite{huang2019reconfigurable, di2020smart}) and assumes negligible amplitude variation and inter-element coupling.
The channel contribution via the $j$-th \gls{EMS} is given by
\begin{equation}
    \mathbf{h}_j^{\mathrm{EMS}}(\boldsymbol{\Phi}_j) = \mathbf{H}_j^{\mathrm{o}} \, \boldsymbol{\Phi}_j \, \mathbf{h}_j^{\mathrm{i}},
    \label{eq:ems_channel}
\end{equation}
where
\begin{itemize}
    \item $\mathbf{h}_j^{\mathrm{i}} \in \mathbb{C}^{L \times 1}$ is the channel vector from the \gls{ue} to the $L$ elements of \gls{EMS} $j$, where the index $i$ stands for the incident wave.
    \item $\mathbf{H}_j^{\mathrm{o}} \in \mathbb{C}^{N_{\mathrm{BS}} \times L}$ is the channel matrix from the $L$ elements of \gls{EMS} $j$ to the $N_{\mathrm{BS}}$ antennas at the \gls{BS}, where the index $O$ stands for the reflection wave.
\end{itemize}

\subsection{Channel Model}
\label{subsec:channel_model}
We assume a block-fading channel with independent fading for each link. In this context, both the direct and EMS-assisted channels are modeled as deterministic multipath propagation, using ray tracing tools. Specifically, we employ the open-source Sionna Ray Tracing engine~\cite{SionnaRT}, which allows for detailed electromagnetic simulation in realistic environments.

The geometric scenario, including the positions and geometries of the \gls{BS}, \gls{ue}, \gls{EMS}s, and relevant scatterers, is built using Blender, an open-source 3D modeling tool. This 3D environment is imported into Sionna RT~\cite{SionnaRT}, which simulates the propagation environment and outputs, for each link, a set of $P$ deterministic multipath components. Each path $p$ is characterized by a complex gain $\alpha_p$, departure and arrival angles $\boldsymbol{\vartheta}^p = (\theta^p, \varphi^p)$, path length, and delay.

The resulting channel impulse response is constructed as
\begin{equation}
    \mathbf{h} = \frac{1}{\sqrt{P}} \sum_{p=1}^{P} \alpha_p \, \varrho(\boldsymbol{\vartheta}^p) \, \mathbf{a}(\boldsymbol{\vartheta}^p),
    \label{eq:generic_channel}
\end{equation}
where $\varrho(\boldsymbol{\vartheta}^p)$ is the element radiation pattern, modeled as in~\cite{3GPP} for \gls{BS} antennas and as in~\cite{CIRS} for \gls{EMS} meta-atoms, and $\mathbf{a}(\boldsymbol{\vartheta}^p) \in \mathbb{C}^{N \times 1}$ is the array response vector, with $N = N_{\mathrm{BS}}$ for the \gls{BS} and $N = L$ for the \gls{EMS}.

The array response vector is given by
\begin{equation}
    \mathbf{a}(\boldsymbol{\vartheta}) = \left[e^{j \mathbf{k}(\boldsymbol{\vartheta})^\top \mathbf{p}_1}, \ldots, e^{j \mathbf{k}(\boldsymbol{\vartheta})^\top \mathbf{p}_N}\right]^\top,
    \label{eq:array_response}
\end{equation}
where $\mathbf{k}(\boldsymbol{\vartheta}) \in \mathbb{R}^{3 \times 1}$ is the wave vector,
\begin{equation} \label{eq:wavevector}
    \mathbf{k}(\boldsymbol{\vartheta}) = \frac{2\pi}{\lambda} 
    \begin{bmatrix}
        \cos(\varphi)\cos(\theta) \\
        \cos(\varphi)\sin(\theta) \\
        \sin(\varphi)
    \end{bmatrix},
\end{equation}
and $\mathbf{p}_n \in \mathbb{R}^{3}$ is the global position of the $n$-th antenna or meta-atom.

All coordinates, including those for the \gls{BS}, \gls{ue}, and \gls{EMS} elements, are referenced in the same global coordinate system for unambiguous modeling.

\textbf{Remark:} We adopt a narrowband model in which the excess delays of the dominant paths are smaller than the pulse width, so echoes overlap within one symbol and are not resolved as separate taps. The received channel is therefore a coherent superposition of paths, where each path delay enters as a carrier-phase rotation. We deliberately use a 10\,MHz bandwidth so that most echoes produced by the considered geometry (including multi-bounce components) fall within the pulse width and are aggregated (with their \emph{true} accumulated phases), rather than being time-resolved as in a wideband model. Moreover, this choice decreases the effective noise power.
\section{Channel Charting Methods}\label{sec:cc_method}

\gls{cc} aims to learn a low-dimensional representation of the spatial relationships between channel states, leveraging features derived from \gls{csi}. To enable both training and evaluation, we consider a set of \gls{TP}s $\mathcal{U} = \{\mathbf{p}_1, \ldots, \mathbf{p}_{N_\mathcal{U}}\}$, where each $\mathbf{p}_u \in \mathbb{R}^3$ denotes a possible \gls{ue} location. Not all \gls{TP}s correspond to active \gls{ue}s at any given time, but \gls{csi} is collected for each.

\subsection{\gls{csi} Feature Construction and Dissimilarity Metrics}

A central step in \gls{cc} is the extraction of features that capture the distinguishing spatial characteristics of the channel. In this work, we adopt the channel covariance matrix as the \gls{csi} feature. The covariance, a large-scale statistic, evolves slowly with position and can be estimated robustly in practice~\cite{kazemi2023beam}. It incorporates both direct and \gls{EMS}-assisted multipath contributions.

For each test point $u$, we compute the covariance matrix as
\begin{equation}\label{eq:cov_matrix}
\mathbf{R}_u(\mathcal{S}) = \mathbb{E}\left[\mathbf{h}_u(\mathcal{S}) \mathbf{h}_u^H(\mathcal{S})\right],
\end{equation}
where the expectation is taken over fading, multipath, and estimation errors (modeled as \gls{SNR}-dependent noise~\cite{STMM}). Here, $\mathcal{S}$ denotes the \gls{EMS} configuration.

To quantify the dissimilarity between two channel states, we employ the \gls{LE} distance between covariance matrices, which is effective for comparing high-dimensional Hermitian matrices:
\begin{align}
    d^{\mathrm{LE}}_{u,u'}(\mathcal{S}) &= \|\log \mathbf{R}_u(\mathcal{S}) - \log \mathbf{R}_{u'}(\mathcal{S})\|_F \\ \notag
    &= \sqrt{\operatorname{Tr}(\boldsymbol{\Lambda}(\mathcal{S}) \boldsymbol{\Lambda}^H(\mathcal{S}))},
\end{align}
where $\log(\cdot)$ denotes the matrix logarithm (computed via SVD~\cite{kazemi2023beam}), and $\boldsymbol{\Lambda}(\mathcal{S}) = \log \mathbf{R}_u(\mathcal{S}) - \log \mathbf{R}_{u'}(\mathcal{S})$. This dissimilarity metric forms the basis of the channel chart.

\subsection{Nonlinear Dimensionality Reduction: t-Distributed
Stochastic Neighbor Embedding }

To embed the dissimilarity structure into a low-dimensional chart, we use \gls{tsne}~\cite{vanDerMaaten2008TSNE}. \gls{tsne} operates by matching the probability distributions of pairwise similarities in the high-dimensional feature space and the low-dimensional latent space.

Let $\mathbf{D}(\mathcal{S}) \in \mathbb{R}^{N_\mathcal{U}\times N_\mathcal{U}}$ be the matrix of \gls{LE} distances. For each $u$, similarities to all other points are defined via a Gaussian kernel:
\begin{equation}
p_{u|u'}(\mathcal{S}) = \frac{\exp\left(-[\mathbf{D}(\mathcal{S})]_{u,u'}^2/2\sigma_u^2\right)}{\sum_{w \neq u} \exp\left(-[\mathbf{D}(\mathcal{S})]_{u,w}^2/2\sigma_u^2\right)},
\end{equation}
with $\sigma_u$ chosen such that the conditional probability distribution $p_{u|u'}$ achieves a specified \emph{perplexity}, a user-selected parameter that determines the effective number of nearest neighbors considered for each point and thus balances the preservation of local and global structure. 
%
%
%
%

In the low-dimensional latent space, we seek an embedding $\mathcal{Z} = \{\mathbf{z}_u\}_{u=1}^{N_\mathcal{U}} \subset \mathbb{R}^{d_{\text{lat}}}$, where each $\mathbf{z}_u$ is the image of test point $u$ in the latent space of dimension $d_{\text{lat}}$ (typically $d_{\text{lat}}=2$ or $3$). To quantify the similarity between pairs of latent points $(u, u')$, \gls{tsne} employs a heavy-tailed Student-$t$ distribution (with one degree of freedom) centered at each point. This kernel allows the model to assign relatively high similarity to points that are moderately distant in the embedding, which helps alleviate the so-called \emph{crowding problem}\footnote{A phenomenon where high-dimensional data cannot be faithfully represented in a lower-dimensional space without severe compression of pairwise distances~\cite{vanDerMaaten2008TSNE}.}. By using the Student-$t$ kernel, \gls{tsne} effectively allocates more area in the latent space to represent moderate and large pairwise distances, thus preserving both local and some global data structure.

Specifically, the similarity between latent points $u$ and $u'$ is defined as:
\begin{equation}
    q_{u,u'} = \frac{\left(1 + \|\mathbf{z}_u - \mathbf{z}_{u'}\|^2\right)^{-1}}{\displaystyle\sum_{w \neq v} \left(1 + \|\mathbf{z}_w - \mathbf{z}_v\|^2\right)^{-1}},
\end{equation}
where the numerator assigns higher similarity to closer points, and the denominator normalizes the values over all distinct pairs $(w, v)$ in the dataset.

The objective of \gls{tsne} is to arrange the latent points $\{\mathbf{z}_u\}$ such that the distribution of pairwise similarities $Q = \{q_{u,u'}\}$ in the latent space matches as closely as possible the target similarity distribution $P = \{p_{u,u'}\}$ derived from the high-dimensional feature space. This is formalized as the minimization of the Kullback–Leibler (KL) divergence from $P$ to $Q$:
\begin{equation}
    \hat{\mathcal{Z}}(\mathcal{S}) = \arg\min_{\mathcal{Z}} \sum_{u,u'} p_{u,u'}(\mathcal{S}) \log \frac{p_{u,u'}(\mathcal{S})}{q_{u,u'}},
\end{equation}
where $p_{u,u'}(\mathcal{S})$ are the joint probabilities from the primary (feature) space, and $q_{u,u'}$ are those in the latent space.

This objective is optimized using gradient descent. The gradient of the KL divergence for a single latent coordinate $\mathbf{z}_u$ is given by:
\begin{equation}
    \frac{\partial f_{\text{t-SNE}}}{\partial\mathbf{z}_u} =
     4\sum_{u'} \left(p_{u,u'}(\mathcal{S}) - q_{u,u'}\right)
     \frac{\mathbf{z}_u - \mathbf{z}_{u'}}{1 + \|\mathbf{z}_u - \mathbf{z}_{u'}\|^2}.
\end{equation}
This gradient forces latent points to move closer together when their similarity in the primary space is underrepresented in the latent space, and to move apart when their similarity is overrepresented. Optimization proceeds by iteratively updating the latent coordinates $\mathcal{Z}$ to minimize the divergence, typically using momentum and early exaggeration strategies to accelerate and stabilize convergence~\cite{vanDerMaaten2008TSNE}.

Standard \gls{tsne} is unsupervised, meaning the latent coordinates $\{\mathbf{z}_u\}$ are determined solely by the structure of the input dissimilarity matrix and have no direct connection to real-world coordinates. To enable actual localization, we adopt a  \gls{stsne}~\cite{zhang2021semi}, in which the latent positions of a subset of labeled points $\mathcal{I} \subset \mathcal{U}$ are “clamped” to their known physical coordinates $\{\mathbf{y}_i\}_{i \in \mathcal{I}}$ throughout the optimization process. This acts as a set of anchor points, guiding the remaining (unlabeled) embeddings to align with the true spatial geometry, while still preserving the local and global structure imposed by the dissimilarities. 

During each iteration, only the unlabeled latent embeddings $\{\mathbf{z}_u : u \notin \mathcal{I}\}$ are updated via gradient descent, while the labeled points remain fixed. Momentum and early exaggeration are applied as in~\cite{vanDerMaaten2008TSNE} to accelerate convergence and improve the fidelity of local neighborhoods.

\begin{algorithm}[t]
\SetAlgoLined
\KwData{Dissimilarity matrix $\mathbf{D}(\mathcal{S})$; labeled index set $\mathcal{I}$; ground-truth coordinates $\{\mathbf{y}_i\}_{i \in \mathcal{I}}$; number of iterations $T$; learning rate $\xi$; momentum parameter $\beta$; exaggeration factor $\gamma$; initial embeddings $\{\mathbf{z}_u^{(0)}\}$}
\KwResult{Optimized latent embeddings $\{\mathbf{z}_u^{(T)}\}$}
\BlankLine
\textbf{Initialization:}
\begin{itemize}
  \item Set $\mathbf{z}_i^{(0)} = \mathbf{y}_i$ for all $i \in \mathcal{I}$ (clamp labeled points)
  \item Randomly initialize $\mathbf{z}_u^{(0)}$ for $u \notin \mathcal{I}$
  \item Set $\mathbf{z}_u^{(-1)} = \mathbf{z}_u^{(0)}$ for all $u$
\end{itemize}

\For{$t = 1$ \KwTo $T$}{
  \For{all $u, u'$}{
    Compute joint probabilities $p_{u,u'}(\mathcal{S}) = \frac{1}{2}(p_{u|u'}(\mathcal{S})+p_{u'|u}(\mathcal{S}))$, with exaggeration factor $\gamma$ applied to $p_{u,u'}$ during early iterations
  }
  \For{all $u, u'$}{
    Compute $q_{u,u'}$ via the latent Student-$t$ kernel
  }
  \For{each $u \notin \mathcal{I}$}{
    Compute gradient $\Delta_u^{(t)} = \frac{\partial f_{\text{t-SNE}}}{\partial \mathbf{z}_u}$
    \;
    Update embedding:
    \[
      \mathbf{z}_u^{(t)} = \mathbf{z}_u^{(t-1)} + \xi \Delta_u^{(t)} + \beta \left( \mathbf{z}_u^{(t-1)} - \mathbf{z}_u^{(t-2)} \right)
    \]
  }
  \For{each $i \in \mathcal{I}$}{
    \textbf{Clamp:} Set $\mathbf{z}_i^{(t)} = \mathbf{y}_i$
  }
}
\caption{\gls{stsne} for Channel Charting and Localization}
\label{alg:stsne}
\end{algorithm}

The early exaggeration factor $\gamma$ is typically used for the first several hundred iterations to amplify attractive forces between points, thereby improving the preservation of local neighborhoods and accelerating convergence, as proposed in~\cite{vanDerMaaten2008TSNE}.

\subsection{Nonlinear Dimensionality Reduction: Fully Connected Autoencoder}\label{sec:AE}
A \gls{FCAE} offers a parametric mapping from high-dimensional \gls{csi} features to a low-dimensional latent chart and back. For each $u \in \mathcal{U}$, we use the standardized, vectorized log-covariance as input:
\begin{equation}
  \widehat{\mathbf{u}}_u(\mathcal{S}) =
    \frac{\mathbf{u}_u(\mathcal{S}) - \boldsymbol{\mu}}{\boldsymbol{\sigma}},
\end{equation}
where $\mathbf{u}_u$ stacks real and imaginary parts of the vectorized upper triangle of $\widetilde{\mathbf{R}}_u = \log(\mathbf{R}_u)$, and $\boldsymbol{\mu}$, $\boldsymbol{\sigma}$ are computed on the training set.

\begin{figure*}[t]
  \centering
  \includegraphics[width=0.8\linewidth]{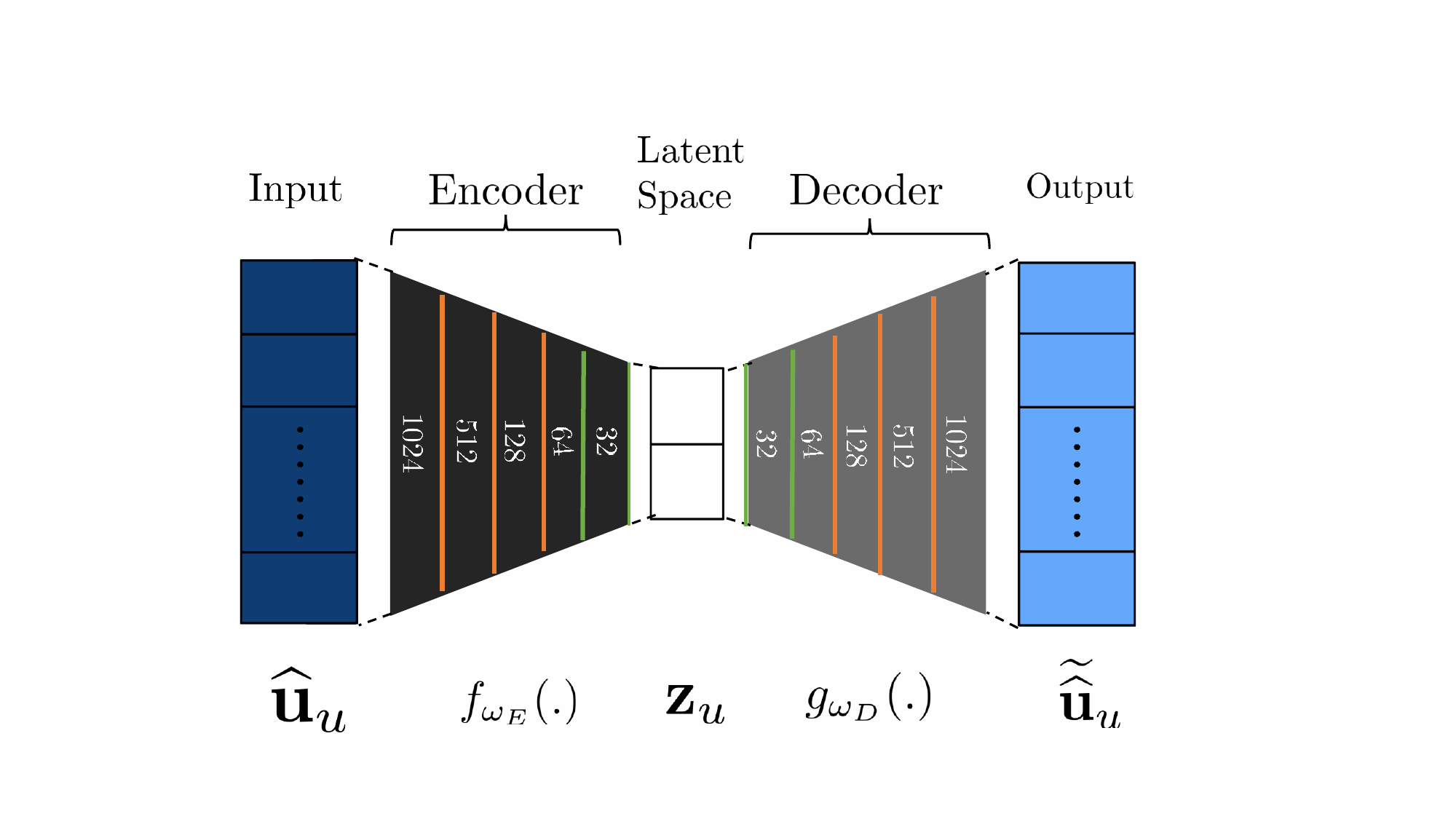}
  \caption{Structure of the \gls{FCAE} used for channel charting. The orange and green colors represent the ReLU and tanh activation functions, respectively, used in the hidden layers.}
  \label{fig:topologyFCAE}
\end{figure*}

The encoder $f_{\omega_E}$ maps $\widehat{\mathbf{u}}_u$ to latent $\mathbf{z}_u \in \mathbb{R}^{d_{\text{lat}}}$, and the decoder $g_{\omega_D}$ reconstructs $\hat{\mathbf{u}}_u$. Both are multilayer neural networks (see Fig.~\ref{fig:topologyFCAE}).

The semi-supervised training loss is
\begin{align}
  L_{\text{AE}} &= \frac{1}{|\mathcal{U}|}\sum_{u} \|\widehat{\mathbf{u}}_u - g_{\omega_D}(f_{\omega_E}(\widehat{\mathbf{u}}_u))\|_2^2 + \frac{\eta}{2}\|\omega_{\!FC}\|_2^2, \\
  L_{\text{E}} &= \frac{1}{|\mathcal{L}|}\sum_{u\in\mathcal{L}} \|\mathbf{z}_u - \mathbf{y}_u\|_2^2 + \frac{\eta}{2}\|\omega_E\|_2^2, \\
  L_{\text{D}} &= \frac{1}{|\mathcal{L}|}\sum_{u\in\mathcal{L}} \|\widehat{\mathbf{u}}_u - g_{\omega_D}(\mathbf{y}_u)\|_2^2,
\end{align}
with $\omega_{\!FC} = \{\omega_E, \omega_D\}$ and $\eta$ a regularization coefficient. The total loss is
\begin{equation}
  L_{\text{tot}} = \alpha L_{\text{AE}} + \beta L_{\text{E}} + \gamma L_{\text{D}}.
\end{equation}

Training proceeds by stochastic gradient descent; see Algorithm~\ref{alg:FCAE}.

\begin{algorithm}[t]
\SetAlgoLined
\KwData{Features $\{\widehat{\mathbf{u}}_u\}$, labels $\{\mathbf{y}_u\}_{u\in\mathcal{L}}$, $\alpha, \beta, \gamma, \eta$, batch size $B$, epochs $T$}
\KwResult{Trained encoder/decoder $(\omega_E, \omega_D)$, latent embeddings $\{\mathbf{z}_u\}$}
\BlankLine
\textbf{Initialize} $\omega_E$, $\omega_D$\;
\For{$t=1$ \KwTo $T$}{
  Sample mini-batch $\mathcal{B} \subset \mathcal{U}$\;
  Compute $L_{\text{AE}}$ on $\mathcal{B}$; compute $L_{\text{E}}, L_{\text{D}}$ on labeled subset\;
  Compute total loss $L_{\text{tot}}$\;
  Update $\omega_E$, $\omega_D$ using gradient of $L_{\text{tot}}$\;
}
\caption{Semi-supervised \gls{FCAE} Training for Channel Charting}
\label{alg:FCAE}
\end{algorithm}

\noindent
The learned latent chart $\{\mathbf{z}_u\}$ is evaluated using metrics in Sec.~\ref{sec:EvaluationMetrics} and used for \gls{EMS} configuration optimization (Sec.~\ref{OptimizationSection}).

\section{Channel Charting in Smart Radio Environment}\label{sec:metrics_opt}

This section develops a rigorous framework for evaluating and optimizing the performance of \gls{cc} in the presence of \gls{EMS}. We begin by defining point-wise metrics—such as localization error, \gls{TW}, and \gls{CT}—which quantitatively assess the fidelity of the learned chart in reflecting true spatial relationships. Next, we present the mathematical parameterization of the \gls{EMS} phase profiles, showing how these profiles directly affect the channel state information and, consequently, the resulting embeddings. We then cast the design of optimal \gls{EMS} phase configurations as a general optimization problem, discussing its non-convex and combinatorial nature and highlighting the key mathematical challenges involved. To address these, we motivate and formalize a codebook-based search approach, which leverages a finite set of physically realizable \gls{EMS} phase patterns to provide a tractable and practical solution.

\subsection{Evaluation Metrics}\label{sec:EvaluationMetrics}
To evaluate the geometric fidelity of the learned chart, three point-wise metrics are considered. 
The first one is the \emph{localization error (LE)}, which quantifies the Euclidean distance between the predicted and actual coordinates of each test point:
\begin{equation}
    \mathrm{LE}_u(\mathcal{S}) = \big\| \hat{\mathbf{z}}_u(\mathcal{S}) - \mathbf{y}_u \big\|_2,
\end{equation}
where $\hat{\mathbf{z}}_u(\mathcal{S})$ denotes the embedded position of user $u$ under configuration $\mathcal{S}$, and $\mathbf{y}_u$ is its true location in the physical domain.

The second metric is \emph{trustworthiness (TW)}, which measures how well the neighborhood relationships in the original feature space are preserved after dimensionality reduction. 
Let $\mathcal{V}_u(\kappa|\mathcal{S})$ be the set of $\kappa$ nearest neighbors of $u$ in the high-dimensional space (computed from the dissimilarity matrix $\mathbf{D}(\mathcal{S})$), and $\mathcal{V}'_u(\kappa|\mathcal{S})$ the corresponding neighbors in the latent space. 
The TW score is then given by
\begin{equation}
\mathrm{TW}_u(\kappa|\mathcal{S}) = 1 - \eta \sum_{\substack{u' \notin \mathcal{V}_u(\kappa|\mathcal{S}) \\ u' \in \mathcal{V}'_u(\kappa|\mathcal{S})}} \big( r'_{u,u'}(\mathcal{S}) - \kappa \big),
\end{equation}
where $r'_{u,u'}(\mathcal{S})$ is the rank of $u'$ in the latent-space neighbor list, and
\begin{equation}
    \eta = \frac{2}{\kappa(2|\mathcal{L}'| - 3\kappa - 1)}.
\end{equation}
Values of TW close to one indicate better preservation of local geometry.

Finally, \emph{continuity (CT)} assesses how well nearby points in the original space remain close in the latent domain. It is computed as
\begin{equation}
\mathrm{CT}_u(\kappa|\mathcal{S}) = 1 - \eta \sum_{\substack{u' \in \mathcal{V}_u(\kappa|\mathcal{S}) \\ u' \notin \mathcal{V}'_u(\kappa|\mathcal{S})}} \big( r_{u,u'}(\mathcal{S}) - \kappa \big),
\end{equation}
where larger CT values (closer to one) correspond to better continuity across neighborhoods.

For any of the above metrics, denoted generically as $m_u(\mathcal{S})$ (where $m$ represents LE, $-\mathrm{TW}$, or $-\mathrm{CT}$ for minimization), we examine its empirical cumulative distribution $F_m(x|\mathcal{S})$ and corresponding $\alpha$-quantile $Q_m(\alpha|\mathcal{S})$. 
Rather than minimizing the mean value, the optimization focuses on the upper quantile of $m_u$, which emphasizes the worst-case users—typically those located in challenging \gls{NLoS} regions—and thus provides a more robust design criterion.

\textbf{Remark:} Note that \gls{TW} and \gls{CT} are naturally maximization metrics within [0,1]. Here, we minimize their negated values ($-{\rm TW}$, $-{\rm CT}$) only for notational uniformity, so that all performance metrics can be written under a common minimization formulation. This transformation does not alter the interpretation of the results.

\subsection{\gls{EMS} Phase Profile Parameterization}

Let the incident and desired outgoing wave vectors be
\begin{equation}
  \mathbf{k}_i \triangleq \mathbf{k}(\boldsymbol{\vartheta}_i), \qquad
  \mathbf{k}_o \triangleq \mathbf{k}(\boldsymbol{\vartheta}_o),
\end{equation}
as defined in \eqref{eq:wavevector}. The generalized Snell’s law~\cite{Gutierrez2017,PhysRevApplied.9.034021,CIRS} gives the required tangential phase gradient to achieve the desired reflection:
\begin{equation}
   \mathbf{k}_o - \mathbf{k}_i = \nabla_{\!\!\parallel} \Phi(\mathbf{r}) + \nu(\mathbf{r})\,\mathbf{u}(\mathbf{r}),
\end{equation}
with $\nabla_{\!\!\parallel}$ denoting the tangential gradient, $\Phi(\mathbf{r})$ the phase profile, and $\nu(\mathbf{r})$ a Lagrange multiplier for the normal component. For a planar \gls{EMS}, this reduces to
\begin{equation}
   \Phi(\mathbf{r}) = \Phi_0 + (\mathbf{k}_o - \mathbf{k}_i)^\top \mathbf{r},
\end{equation}
which can be sampled on the discrete \gls{EMS} elements as
\begin{equation}
   \phi_\ell = (\mathbf{k}_o - \mathbf{k}_i)^\top \mathbf{p}_\ell + \Phi_0.
\end{equation}

\subsection{Optimization Problem: Continuous and Codebook-Based Formulation}\label{OptimizationSection}

The goal is to find the \gls{EMS} phase configuration $\mathcal{S}$ that minimizes the $\alpha$-quantile of the LE, negative \gls{TW}, or negative CT evaluated over all test points. Explicitly,
\begin{equation}
    \hat{\mathcal{S}} = \arg\min_{\mathcal{S}\,\in\,\mathbb{S}}\;Q_{{m}}(\alpha|\mathcal{S}),
\end{equation}
where $\mathbb{S}$ is the set of all feasible phase matrices for all \gls{EMS}s. For $M$ \gls{EMS}s of $L$ elements each, $\mathbb{S}$ is the $M \times L$ dimensional torus of elementwise phase shifts:
\begin{equation}
    \mathbb{S} = \left\{ \{\Phi_j\}_{j=1}^M : \Phi_j = \diag(e^{j\boldsymbol{\phi}_j}),\; \boldsymbol{\phi}_j \in [0,2\pi)^L \right\}.
\end{equation}
This problem is high-dimensional, non-convex, and combinatorial, and thus intractable for practical \gls{EMS} sizes. The objective function is highly non-linear in $\mathcal{S}$ due to the complex dependency of the channel and embedding on the \gls{EMS} phase profile, and it have many local minima.

To make optimization tractable and align with \gls{EMS} fabrication constraints, we adopt a \textit{codebook-based approach} \cite{Wu2019, Huang2019, Dai2020, Zhang2022, Tang2020}. Here, a finite codebook $\mathcal{C}$ of $K$ candidate phase profiles is constructed—typically using a range of linear phase gradients or pre-selected angle pairs. The joint codebook for all $M$ \gls{EMS}s is the Cartesian product $\mathbb{C} = \mathcal{C}^M$, with $K^M$ possible configurations:
\begin{equation}
    \hat{\mathcal{S}}
    = \arg\min_{\mathcal{S}\,\in\,\mathbb{C}} Q_{m}(\alpha|\mathcal{S}).
\end{equation}
This approach allows for \textit{exhaustive or greedy search} over a manageable set of profiles, enabling practical design. While it does not guarantee global optimality, it strikes a balance between computational tractability, hardware feasibility, and worst-case performance~\cite{Wu2019,Zhang2022,Dai2020}.

\subsection{Codebook Construction for \gls{EMS} Phase Profiles}

Following the phase-gradient codebook formulation adapted from the previous work ~\cite{Maleki_CC_in_SRE}, the codebook $\mathcal{C}$ is designed to contain a finite number of physically realizable \gls{EMS} phase profiles that enable both practical fabrication and tractable optimization. In this approach, each candidate phase profile represents a linear phase ramp distributed over the \gls{EMS} surface. For a planar \gls{EMS} whose element coordinates are $\mathbf{p}_\ell = (x_\ell, y_\ell)$ with inter-element spacings $d_x$ and $d_y$, we define two sets of quantized phase increments, $\mathcal{C}_x = \{\Delta\phi_x^{(a)}\}_{a=1}^{K_x}$ and $\mathcal{C}_y = \{\Delta\phi_y^{(b)}\}_{b=1}^{K_y}$. Each codeword corresponds to a specific pair of discrete slopes $(a,b)$, and its associated phase profile is expressed as
\begin{equation}
    \phi_\ell^{(a,b)} = \Phi_0^{(a,b)} + \gamma_x^{(a)} x_\ell + \gamma_y^{(b)} y_\ell,
\end{equation}
where $\gamma_x^{(a)} = \Delta\phi_x^{(a)}/d_x$ and $\gamma_y^{(b)} = \Delta\phi_y^{(b)}/d_y$. In the numerical studies presented here, we consider only one-dimensional horizontal phase gradients ($\gamma_y^{(b)} = 0$), so the expression simplifies to
\begin{equation}
    \phi_\ell^{(a)} = \Phi_0^{(a)} + \gamma_x^{(a)} x_\ell.
\end{equation}
Accordingly, the resulting codebook $\mathcal{C}$ comprises $K = K_x$ distinct phase profiles, each associated with a specific steering direction or angular spread.

For each \gls{EMS}, the candidate reflection matrices $\Phi^{(a)} = \mathrm{diag}(e^{j\phi_1^{(a)}}, \dots, e^{j\phi_L^{(a)}})$ are generated in advance. When $M$ \gls{EMS}s are jointly deployed, the corresponding joint search space becomes the Cartesian product $\mathbb{C} = \mathcal{C}^M$.

In practice, the parameters defining the codebook are chosen to span the relevant range of expected propagation directions. The final cardinality $K$ of $\mathcal{C}$ is determined by the balance between design resolution, computational complexity, and the practical limitations of hardware implementation~\cite{Wu2019, Zhang2022}.

\section{Results and Discussion}
\label{sec:sim_results}

This section presents a comprehensive study of the proposed \gls{cc} framework with \gls{EMS}s, in a realistic urban deployment. We provide both quantitative and qualitative evaluations, highlight the influence of codebook-based \gls{EMS} design, and discuss the performance trade-offs observed under various panel configurations.
\begin{figure}[t!]
  \centering
  \begin{subfigure}[b]{0.8\linewidth}
    \centering
    \includegraphics[width=\linewidth]{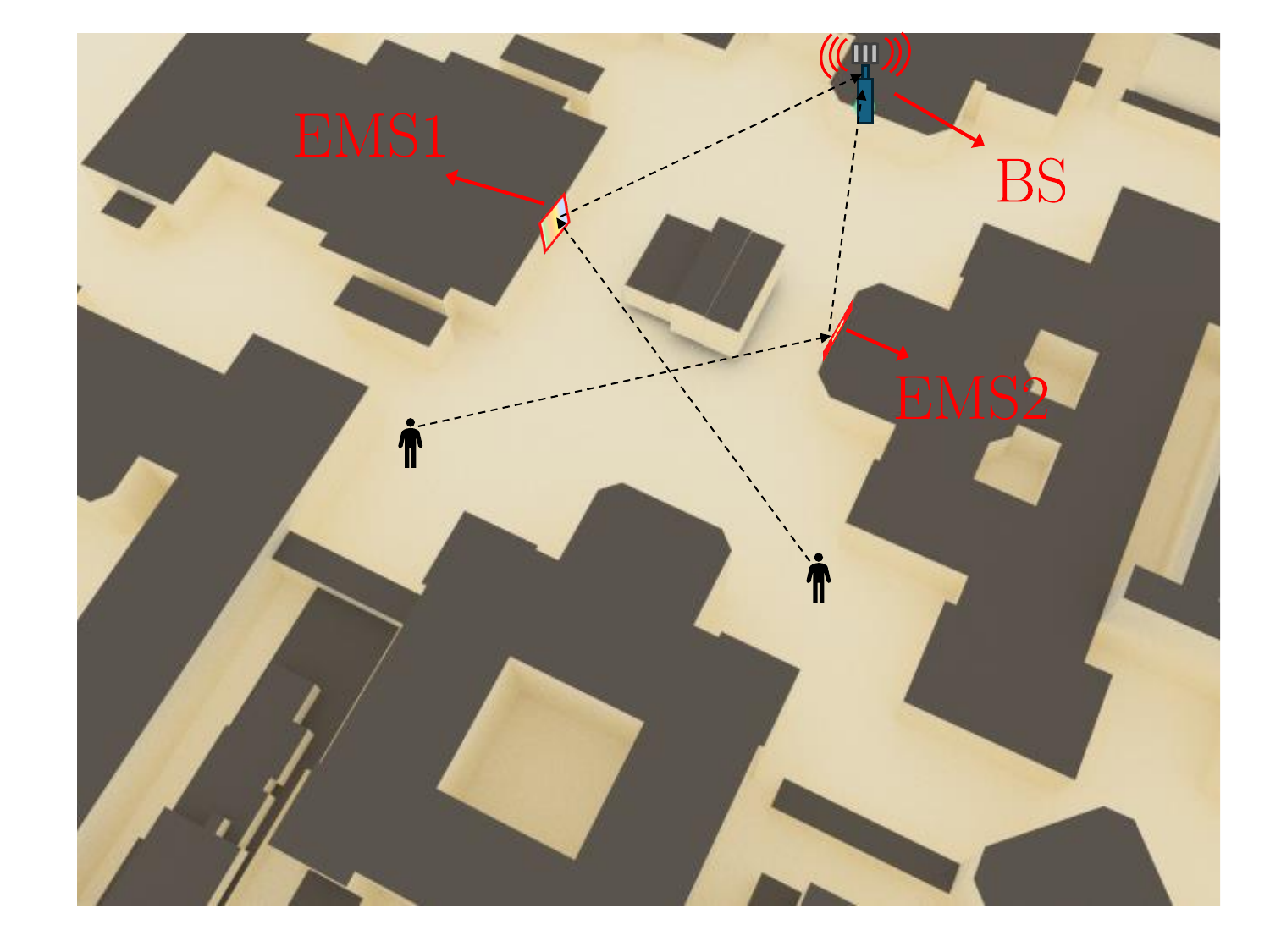}
  \end{subfigure}
  \caption{Geometry of the considered scenario, adapted from \cite{Maleki_CC_in_SRE}}
  \label{fig:topology}
\end{figure}

\subsection{Scenario Description and Simulation Setup}

The scenario, illustrated in Fig.~\ref{fig:topology}, spans an $80 \times 110\,\mathrm{m}^2$ urban area derived from \gls{OSM}. Channels are generated deterministically using the Sionna ray tracer~\cite{hoydis2023sionna}, with path gains ($\alpha_p$) and physical multipath properties accurately reflected in the input \gls{csi} features. The \gls{BS} is placed on top of the tallest building, while two static $60 \times 60$ \gls{EMS}s are mounted on facing building walls at $5.5$~m height.

A total of 3200 \gls{TP}s are uniformly distributed to represent potential user locations, covering both \gls{los} and \gls{nlos} regions. Key simulation parameters are summarized in Table~\ref{tab:var}.

\begin{table}[b!]
    \centering
    \footnotesize
    \caption{Default simulation parameters.}
    \begin{tabular}{l|c|c}
    \toprule
        \textbf{Parameter} &  \textbf{Symbol} & \textbf{Value(s)}\\
        \hline
        Carrier frequency & $f_0$  & $30\,\mathrm{GHz}$ \\
        Bandwidth & $B$ & $10\,\mathrm{MHz}$\\
        \gls{ue} transmit power & $\sigma_s^2$ & $23\,\mathrm{dBm}$\\
        Noise power & $\sigma_n^2$ & $-92\,\mathrm{dBm}$\\
        \gls{EMS} size & $L \times L$ & $60 \times 60$\\
        \gls{EMS} element spacing & $d_n,d_m$ & $\lambda_0/4$ \\
        \gls{BS} antenna array & $N_\mathrm{r}$ & $8\times 4$\\
        \gls{ue} antenna array & $N_\mathrm{t}$ & $1$ \\
        Tx/Rx element spacing & $d_\mathrm{Tx},d_\mathrm{Rx}$ & $\lambda_0/2$ \\
        \gls{BS} height & $h_\mathrm{BS}$ & $8.5\,\mathrm{m}$ \\
        \gls{ue} height & $h_\mathrm{UE}$ & $1.5\,\mathrm{m}$ \\
        \gls{EMS} height & $h_\mathrm{EMS}$ & $5.5\,\mathrm{m}$ \\
        \bottomrule
    \end{tabular}
    \label{tab:var}
\end{table}
\subsection{EMS Phase Codebooks and Methodology}

\begin{figure}[b!]
  \centering
  \includegraphics[width=.49\linewidth]{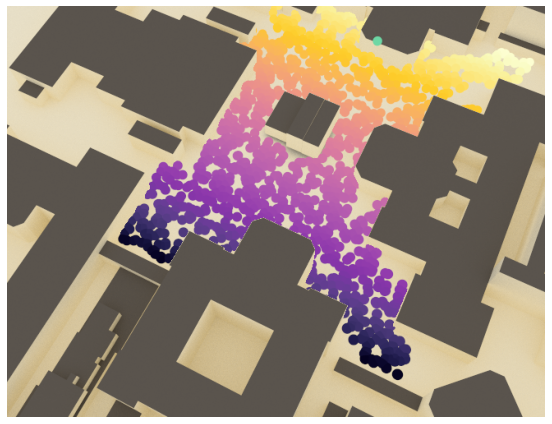}
  \includegraphics[width=.49\linewidth]{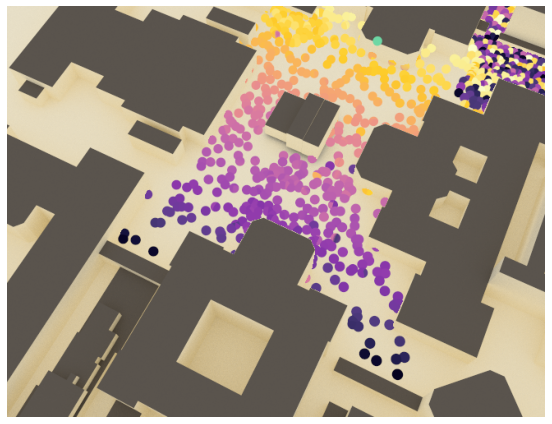}\\
  \includegraphics[width=.49\linewidth]{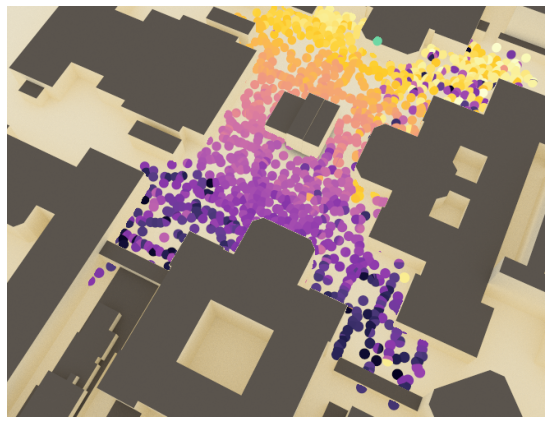}
  \includegraphics[width=.49\linewidth]{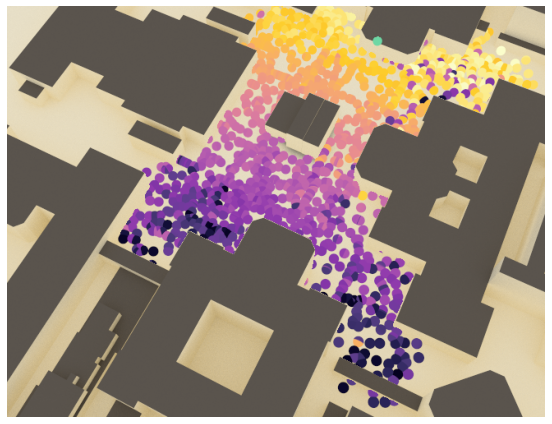}
  \caption{Channel chart embeddings (\gls{stsne}, 15\% supervision): (a) ground truth positions, (b) no \gls{EMS}, (c) specular \gls{EMS}, (d) best codebook \gls{EMS}s. Colors reflect $y$-coordinates. Only codebook-optimized \gls{EMS}s recover full spatial geometry, especially in \gls{nlos}.}
  \label{fig:colorbar}
\end{figure}

\begin{figure}[b!]
  \centering
  \includegraphics[width=.98\linewidth]{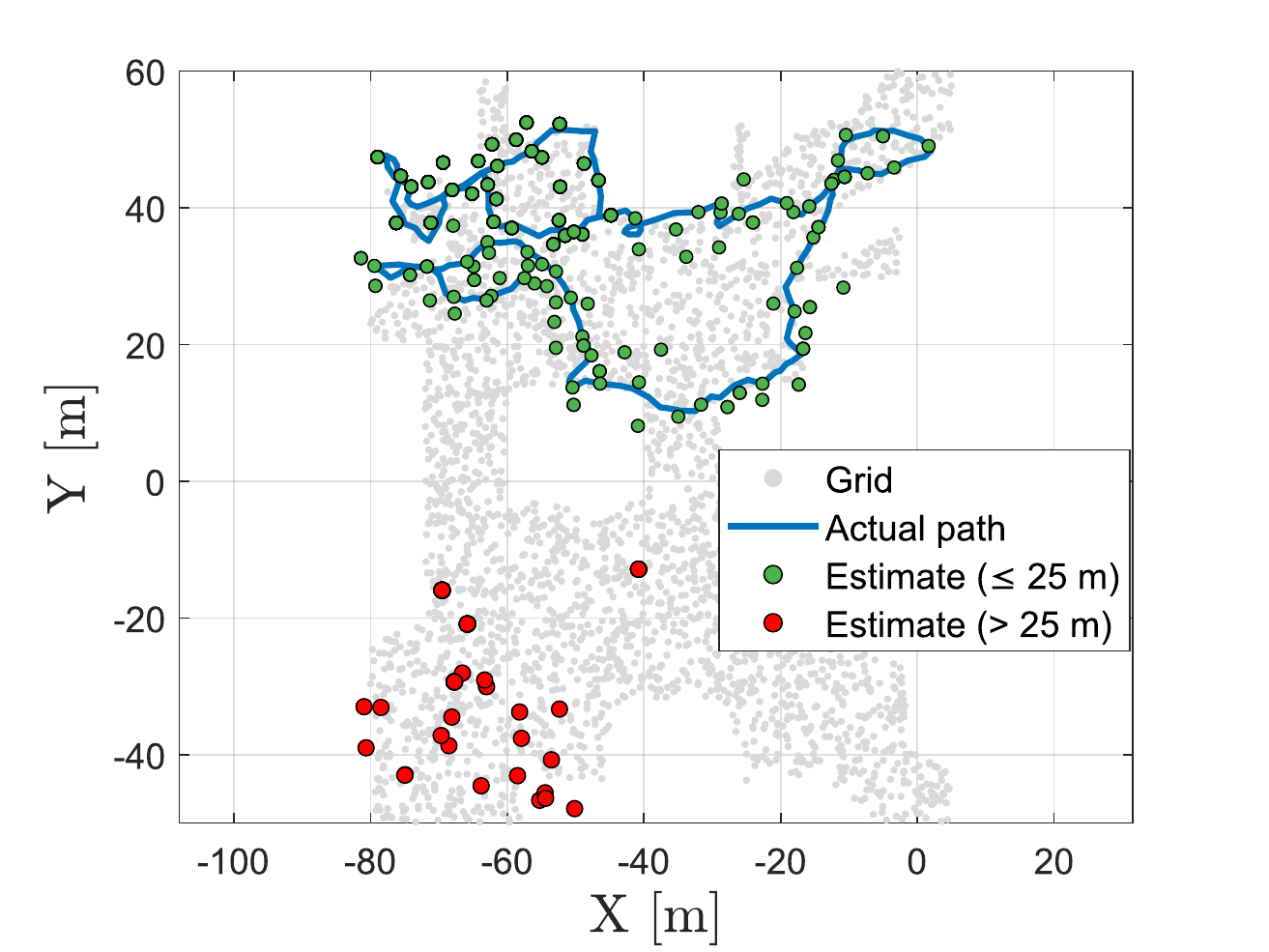}\\
  \includegraphics[width=.98\linewidth]{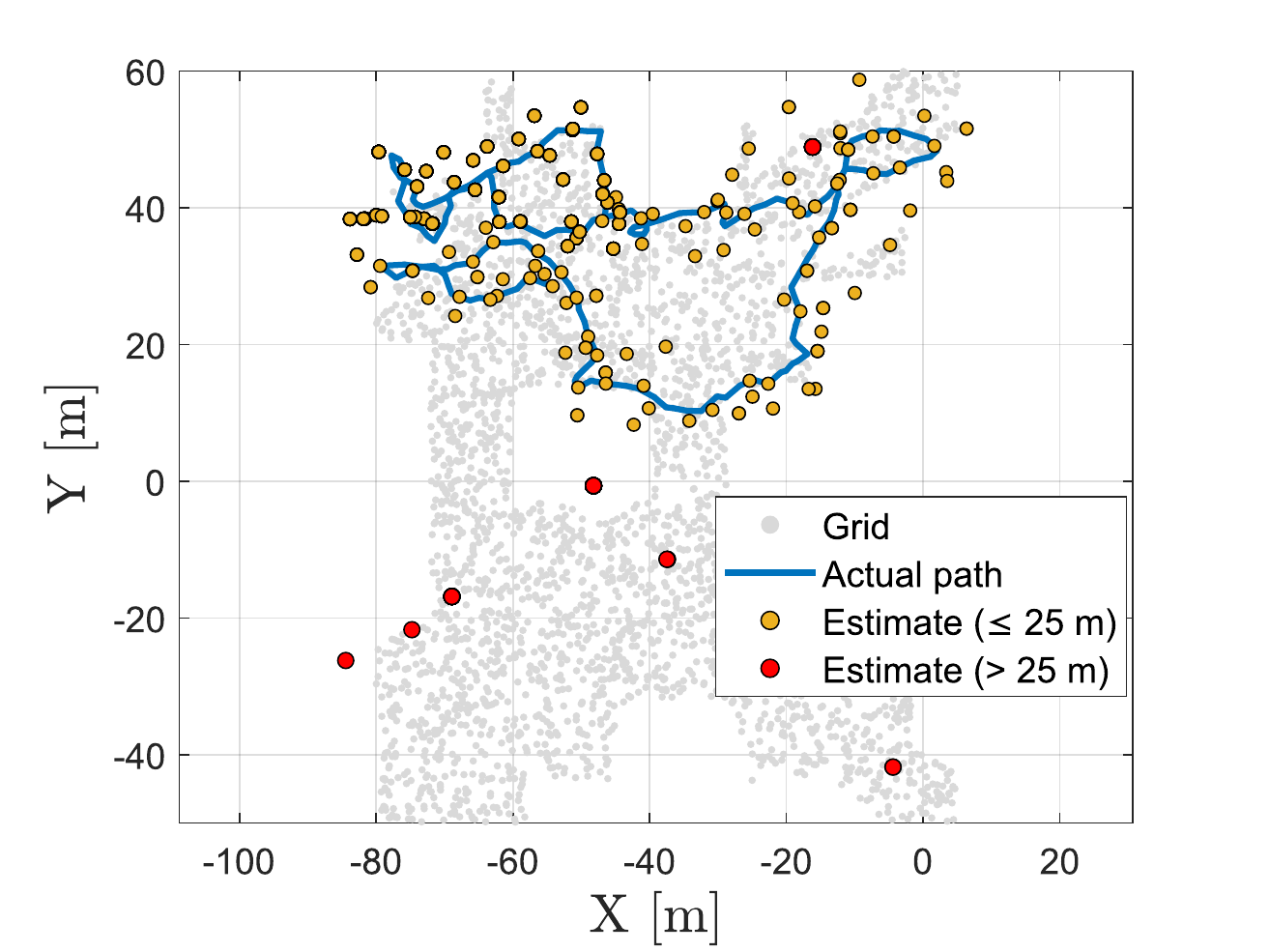}
  \caption{Trajectory estimation using t-SNE (15 \% supervision):
(top) no EMS, (bottom) both EMSs with best codebook.
Blue: ground truth; circles: inferred positions (red = error $>25$ m).
Optimized \glspl{EMS} cut severe outliers by over 4× and yield smooth, gap-free tracking.}
  \label{fig:trajectory}
\end{figure}

For each \gls{EMS}, a codebook of 11 DFT-based horizontal phase gradients is used. The codewords are empirically designed to ensure good coverage of \gls{nlos} regions for this scenario, and the DFT structure guarantees orthogonality between codebook slopes given the \gls{EMS} size. The horizontal orientation is chosen for simplicity and computational tractability. While this is a clear simplification, it is justified by the scale of the optimization problem and aligns with practical \gls{EMS} fabrication limits.

A total of $11\times11=121$ two-panel codeword combinations are thus considered. Increasing the codebook size showed no substantial improvement, confirming that the current discretization is adequate for this environment. Further optimization—such as joint placement and codebook design—would require network planning and is left for future work.

Each method (\gls{tsne} and \gls{ae}) is evaluated with 15\% and 30\% supervision; values below 15\% cause a notable drop in performance, while higher rates do not yield additional gains.

\textbf{Remark 1:} The \gls{EMS} element spacing is set to $\lambda/4$ to satisfy the effective-homogeneity, to ensure that the structure can be homogenized and modeled via \gls{GSTC}. A spacing larger than $\lambda/4$ would break this approximation and is widely recognized as the upper bound for retaining an effective medium description \cite{MTM}.

\textbf{Remark 2:} In all simulations, we assume \gls{csi} is available at the receiver in the form of a noisy estimate of the true channel, rather than perfect knowledge. Specifically, the channel covariance matrices are perturbed according to the received \gls{SNR} at each test point, thereby modeling the impact of estimation errors without relying on an idealized perfect-\gls{csi} assumption.

\subsection{Continuity and Trajectory Tracking}

Figure~\ref{fig:colorbar} provides a qualitative comparison of the embedding quality. The true $y$-coordinate is color-coded, allowing a direct visual check of how well spatial relationships are preserved.

Without \gls{EMS}s, the \gls{nlos} regions collapse into tight clusters in the embedding, destroying the spatial ordering. Specular (mirror-like) panels partially recover the structure, but local errors remain—especially near building edges. Only the best codebook-optimized \gls{EMS}s (selected for 90th-percentile performance) enable a faithful mapping: color bands are uniform and well ordered, with strong separation of distant points even in challenging \gls{NLoS} zones.

Trajectory-based evaluation (Fig.~\ref{fig:trajectory}) highlights the impact on
mobility tracking. Without \glspl{EMS}, the inferred path contains numerous
catastrophic deviations—about 18 \% of trajectory points exceed 25 m error, with
some reaching 70–80 m and breaking continuity. With optimized
codebook-designed \glspl{EMS}, these severe outliers drop to roughly 4 \% of the
points. The trajectory follows the
ground truth smoothly, confirming that static \glspl{EMS} stabilize the channel
chart and greatly improve localization reliability for moving users.

\begin{figure}[t]
  \centering
  \subcaptionbox{\gls{tsne}-based }[.95\linewidth][c]{\includegraphics[width=\linewidth]{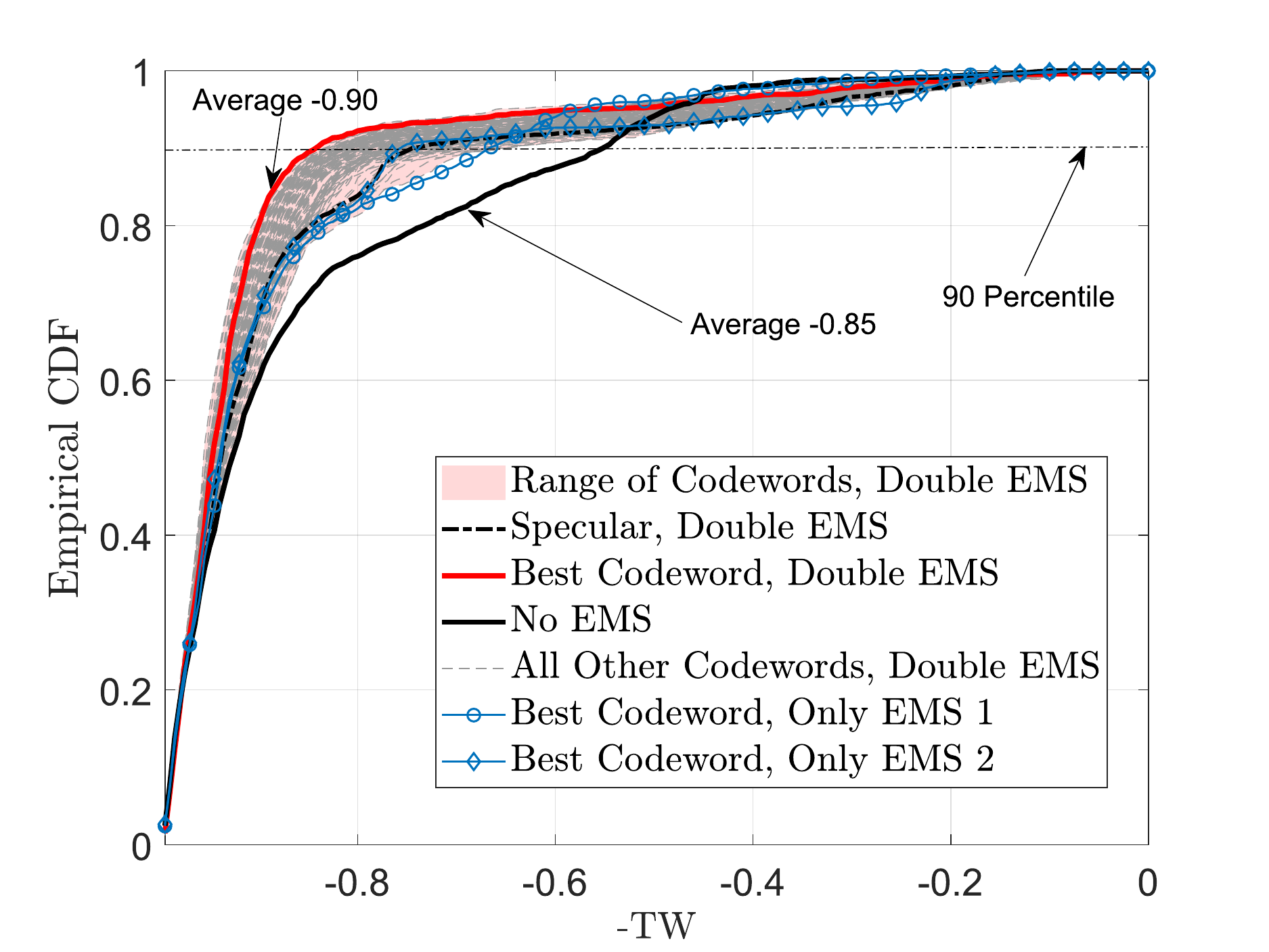}}

  \subcaptionbox{autoencoder-based }[.95\linewidth][c]{\includegraphics[width=\linewidth]{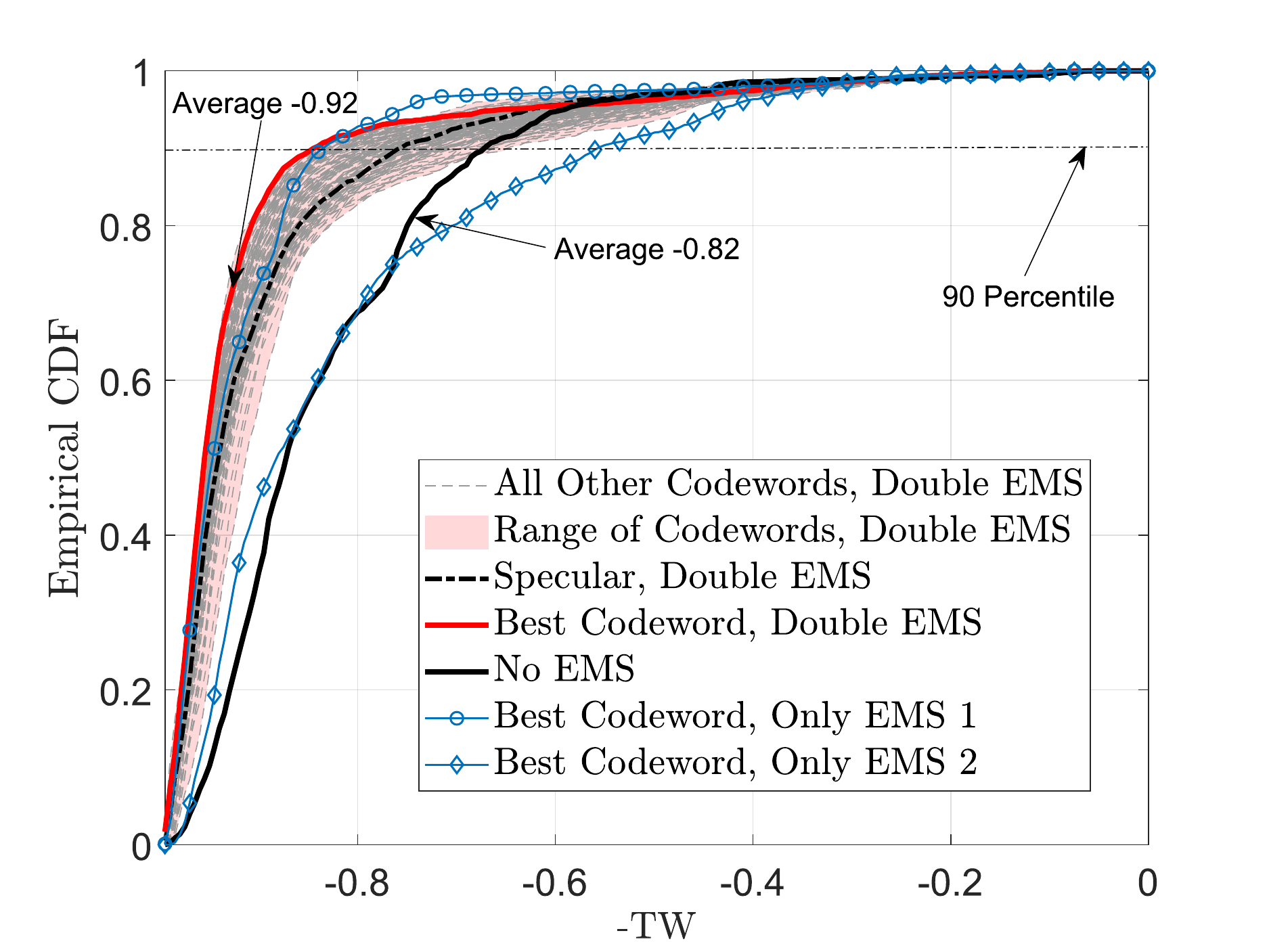}}
  
  \caption{Empirical CDF of $-\mathrm{TW}$ (\gls{tsne}, 15\% supervision): black = no EMS; dash-dotted = specular \gls{EMS}s; colored = best codebook \gls{EMS}s; gray band = all codeword pairs; markers = single-panel.}
  \label{fig:TW}
\end{figure}

\begin{figure}[t]
  \centering
  \subcaptionbox{\gls{tsne}-based }[.95\linewidth][c]{\includegraphics[width=\linewidth]{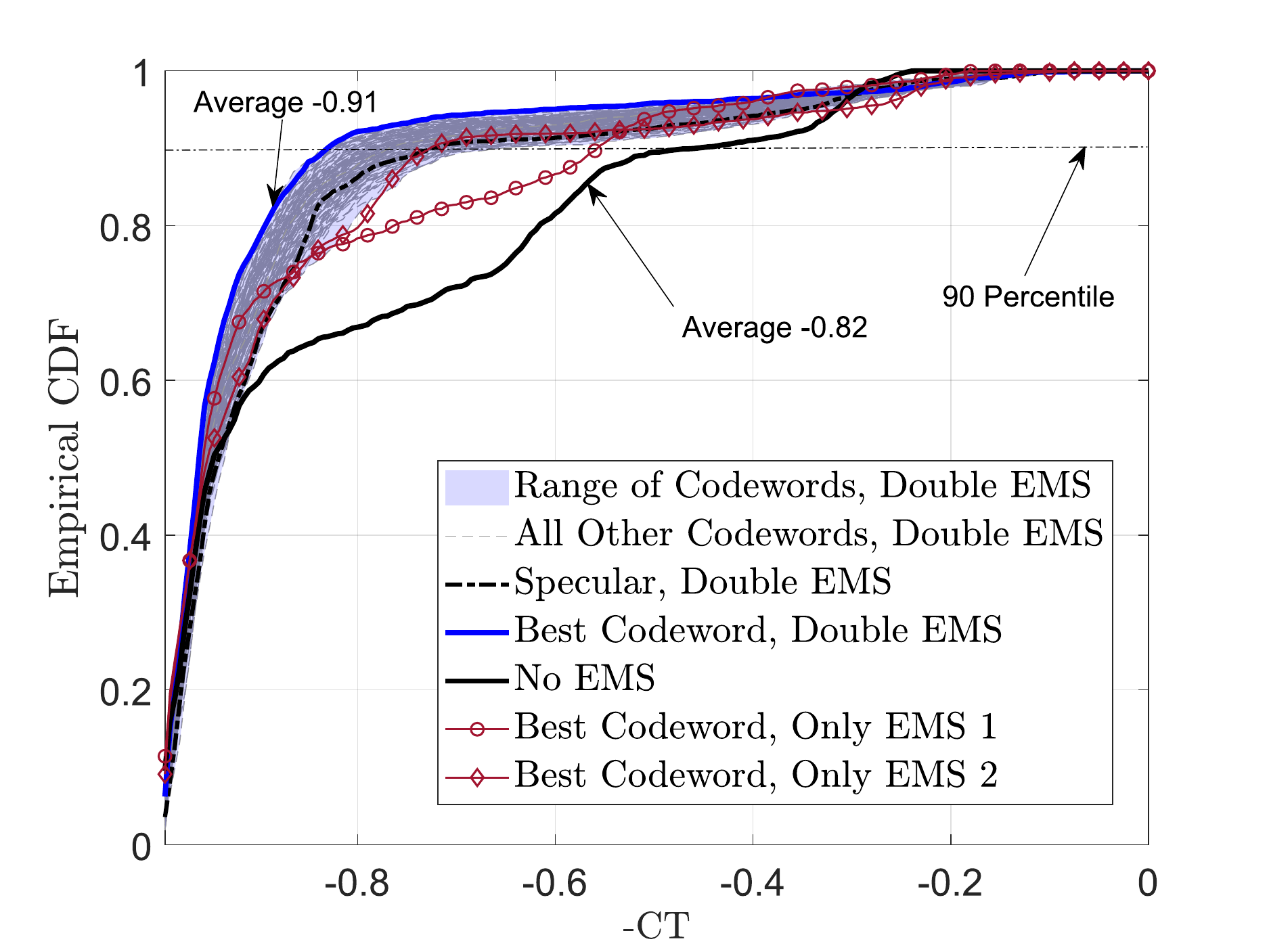}}
  \subcaptionbox{autoencoder-based }[.95\linewidth][c]{\includegraphics[width=\linewidth]{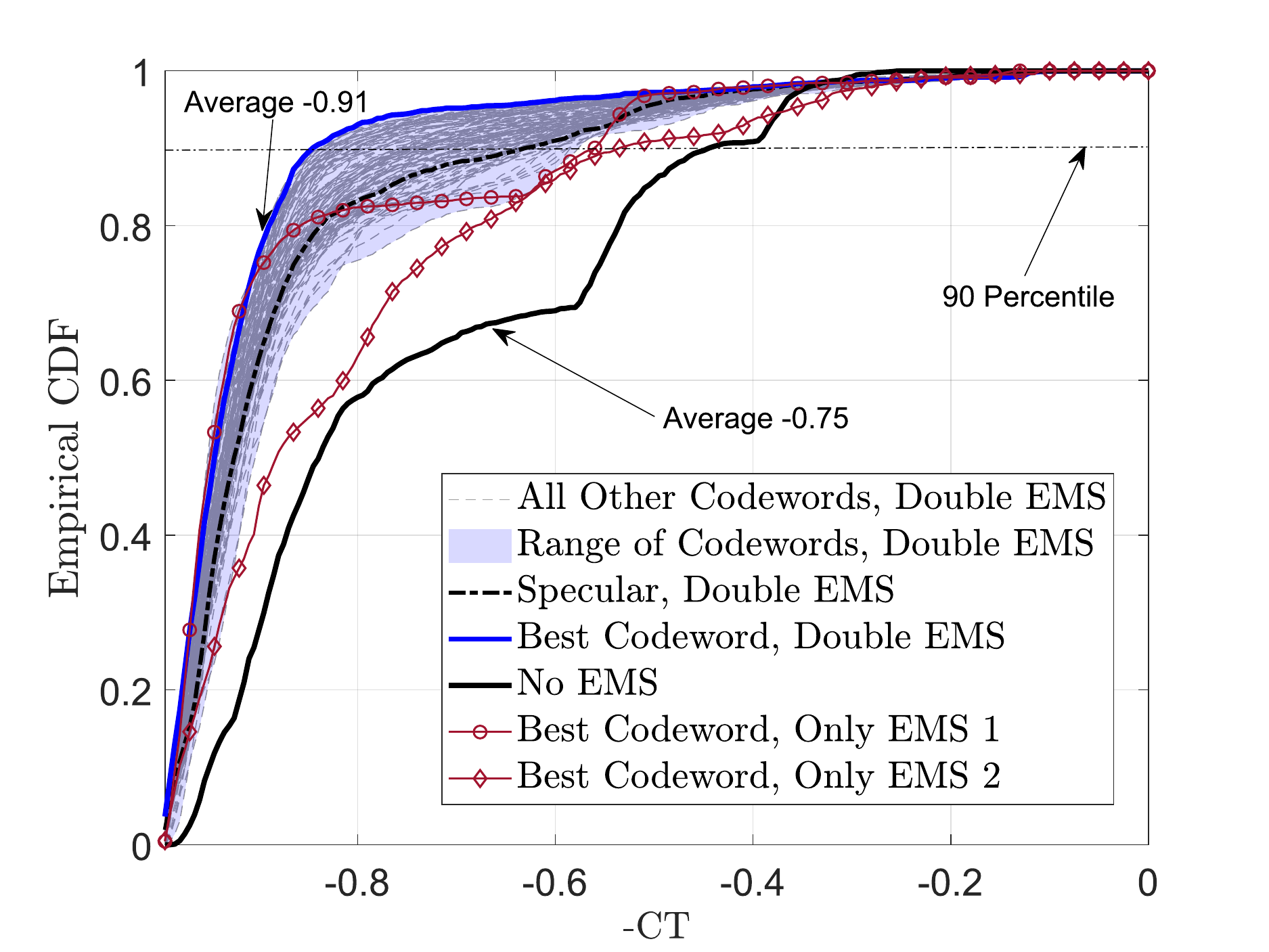}}
  
  \caption{Empirical CDF of $-\mathrm{CT}$ (\gls{tsne}, 15\% supervision).}
  \label{fig:CT}
\end{figure}

\begin{figure}[t]
  \centering
\subcaptionbox{autoencoder-based }[.95\linewidth][c]{ \includegraphics[width=\linewidth]{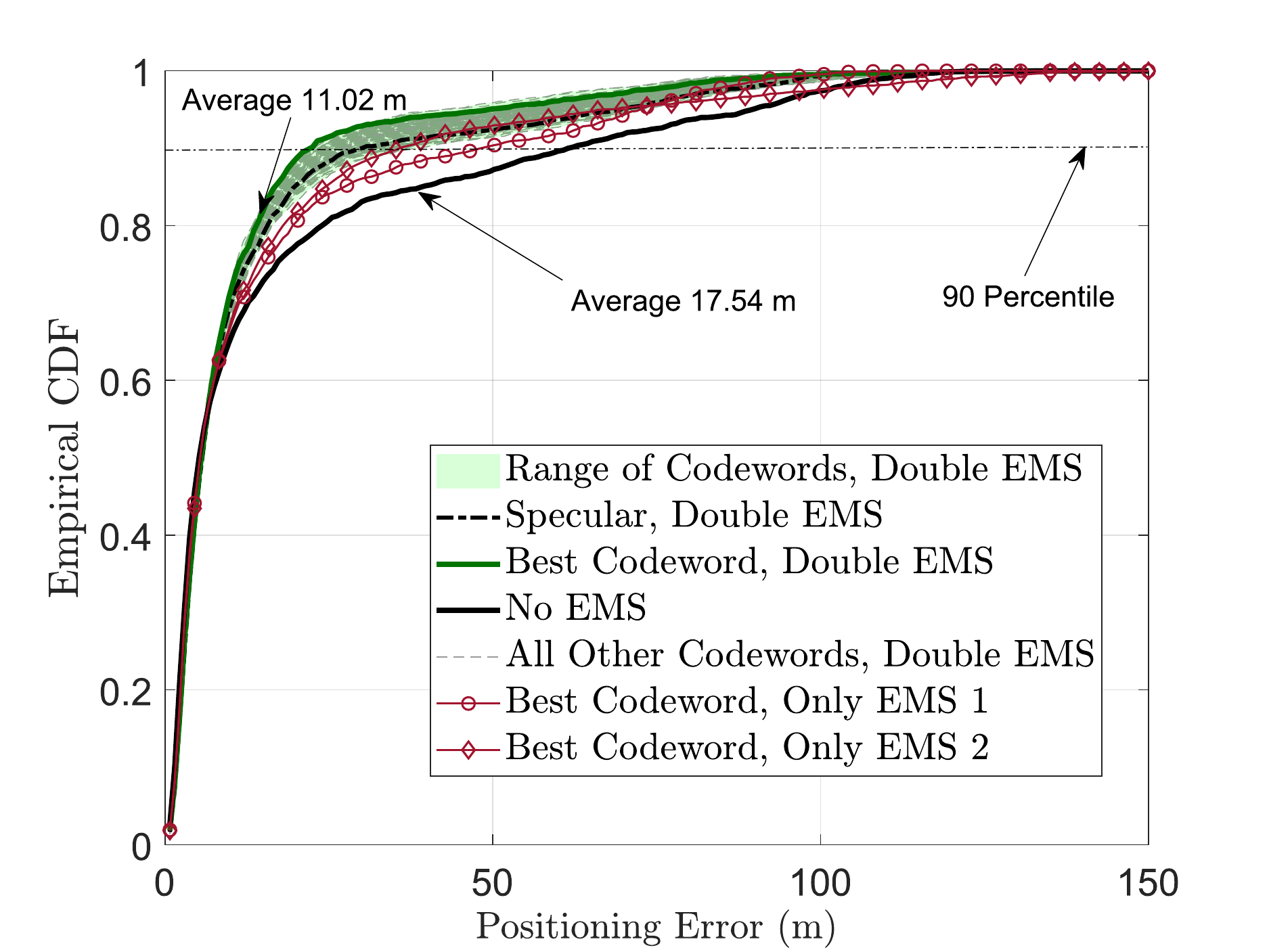}}

\subcaptionbox{autoencoder-based }[.95\linewidth][c]{ \includegraphics[width=\linewidth]{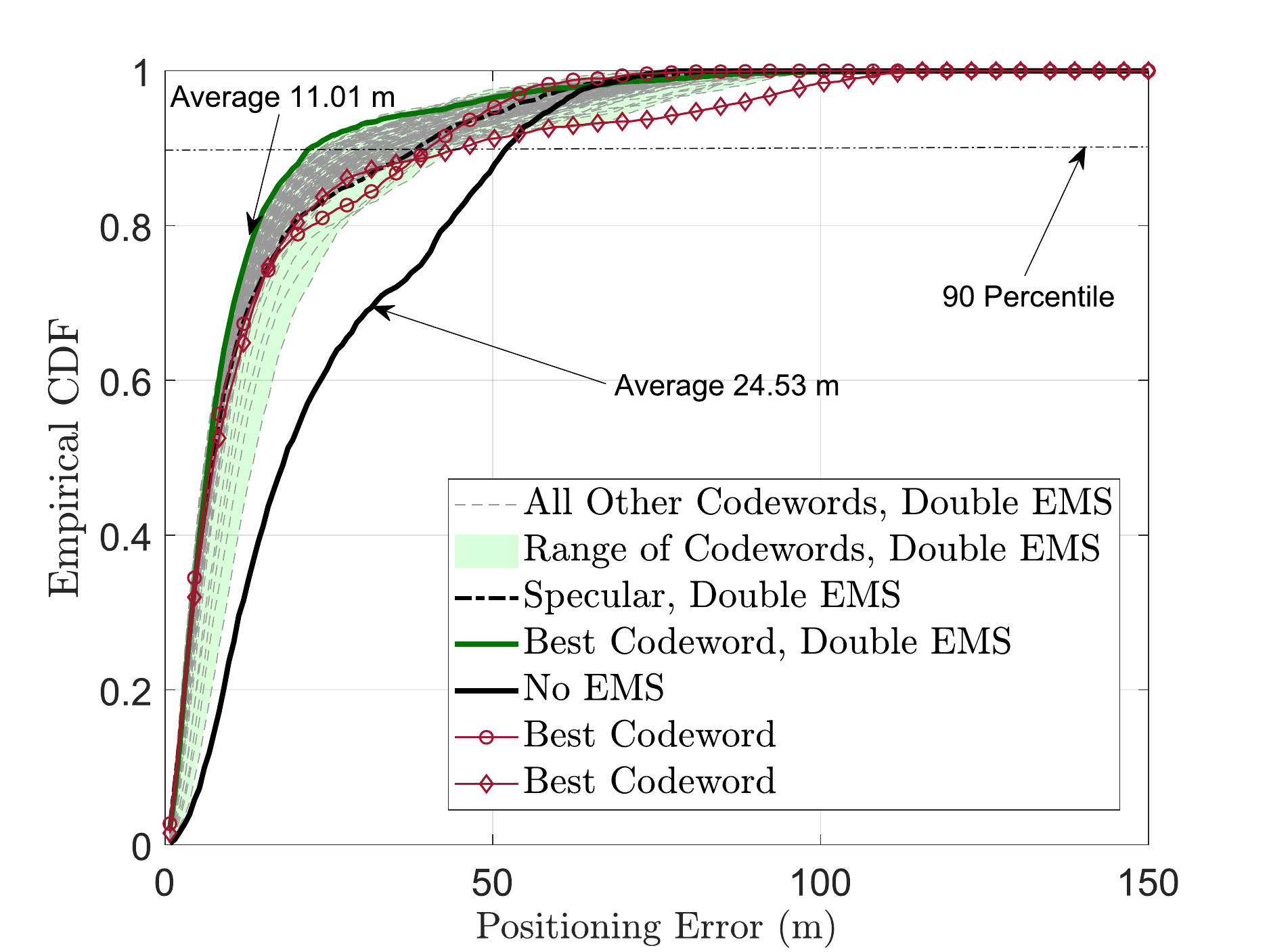}}
\caption{Empirical CDF of positioning error (meters, \gls{tsne}, 15\% supervision).}
  \label{fig:PosError}
\end{figure}

\subsection{Trustworthiness, Continuity, and Positioning Error}

Figures~\ref{fig:TW}--\ref{fig:PosError} show the empirical CDFs for \gls{TW}, \gls{CT}, and positioning error across all test points, for both \gls{tsne} and \gls{ae} methods. The following scenarios are compared:
\begin{itemize}
  \item \textbf{No \gls{EMS}} (solid black): baseline case.
  \item \textbf{Specular \gls{EMS}} (dash-dotted): panels act as mirrors.
  \item \textbf{Best codebook \gls{EMS}s} (solid color): configuration that optimizes the 90th percentile.
  \item \textbf{All other codewords} (gray band): envelope across 121 codeword pairs.
  \item \textbf{Best single-panel} (markers): only \gls{EMS} 1 or \gls{EMS} 2 is active.
\end{itemize}
The x-axis shows the negative of each metric (lower is better); the dashed horizontal line marks the 90th percentile.

Across all metrics, codebook-optimized \gls{EMS}s provide the strongest gains for the 60th to 95th percentiles, i.e., for \gls{nlos} or challenging user positions. Improvements for the lowest-error (best) users are limited, as their paths are already direct. No scenario exhibited a degradation in performance relative to the baseline. This demonstrates that the core benefit of \gls{EMS} optimization lies in enhancing worst-case, not best-case, outcomes. Note that this is scenario-dependent; however, some improvement is always observed when \gls{EMS}s are included.

A summary of mean and 90th-percentile values for each scenario is reported in Table~\ref{tab:metrics_summary}. Codebook-optimized \gls{EMS}s consistently achieve lower errors than both the no-\gls{EMS} baseline and the specular EMS case, with further reductions observed as supervision increases from 15\% to 30\%. An interesting observation is that the \gls{ae} method is much more sensitive to the supervision percentage in an environment without \glspl{EMS}. For instance, increasing the serupervision from 15\% to 30\%, the average localization error with \gls{ae}, drops from 24.53m to 17.12m, while this change is much less for the \gls{tsne} method. However in an \gls{SRE} enabled with two \glspl{EMS}, this severe sensitivity of \gls{ae} to the supervision percentage vanishes.

\begin{table*}[t]
  \centering
  \caption{Summary of evaluation metrics (average and 90th‑percentile) for different \gls{EMS}S scenarios, methods, and supervision levels (15\,\% vs.\ 30\,\%).}
  \label{tab:metrics_summary}
  \begin{tabular}{@{}lllrrrr@{}}
    \toprule
    Metric & Method & Scenario           & \multicolumn{2}{c}{15\,\% Sup.}   & \multicolumn{2}{c}{30\,\% Sup.}   \\
           &        &                    & Avg.     & 90th‑perc.  & Avg.     & 90th‑perc.  \\ 
    \midrule
    \multirow{10}{*}{\shortstack{\gls{CT}\\($-\mathrm{CT}$)}} 
      & \multirow{5}{*}{AE}
        & No EMS           & $-0.75$  & $-0.445$ & $-0.81$      & $-0.43$      \\
      &                      & Specular \gls{EMS}s    & $-0.87$  & $-0.630$ & $-0.88$      & $-0.74$      \\
      &                      & \gls{EMS}1 only (best) & $-0.877$ & $-0.560$ & $-0.88$      & $-0.57$      \\
      &                      & \gls{EMS}2 only (best) & $-0.807$ & $-0.535$ & $-0.87$      & $-0.78$      \\
      &                      & Best double \gls{EMS}  & $-0.91$  & $-0.845$ & $-0.93$      & $-0.88$      \\ 
    \cmidrule(l){2-7}
      & \multirow{5}{*}{\gls{tsne}}
        & No \gls{EMS}           & $-0.82$  & $-0.455$ & $-0.84$      & $-0.53$      \\
      &                      & Specular \gls{EMS}s    & $-0.87$  & $-0.720$ & $-0.89$      & $-0.78$      \\
      &                      & \gls{EMS}1 only (best) & $-0.87$  & $-0.555$ & $-0.89$      & $-0.575$      \\
      &                      & \gls{EMS}2 only (best) & $-0.874$ & $-0.725$ & $-0.888$      & $-0.765$      \\
      &                      & Best double \gls{EMS}  & $-0.91$  & $-0.830$ & $-0.92$      & $-0.88$      \\ 
    \midrule
    \multirow{10}{*}{\shortstack{\gls{TW}\\($-\mathrm{TW}$)}}
      & \multirow{5}{*}{AE}
        & No \gls{EMS}           & $-0.82$  & $-0.67$      & $-0.87$      & $-0.68$      \\
      &                      & Specular \gls{EMS}s    & $-0.89$  & $-0.75$      & $-0.90$      & $-0.74$      \\
      &                      & \gls{EMS}1 only (best) & $-0.91$  & $-0.83$      & $-0.914$      & $-0.83$      \\
      &                      & \gls{EMS}2 only (best) & $-0.82$  & $-0.55$      & $-0.85$      & $-0.67$      \\
      &                      & Best double \gls{EMS}  & $-0.92$  & $-0.84$      & $-0.93$      & $-0.9$      \\ 
    \cmidrule(l){2-7}
      & \multirow{5}{*}{\gls{tsne}}
        & No \gls{EMS}           & $-0.85$  & $-0.55$      & $-0.86$      & $-0.57$      \\
      &                      & Specular \gls{EMS}s    & $-0.87$  & $-0.74$      & $-0.89$      & $-0.81$      \\
      &                      & \gls{EMS}1 only (best) & $-0.88$  & $-0.66$      & $-0.9$      & $-0.68$      \\
      &                      & \gls{EMS}2 only (best) & $-0.87$  & $-0.75$      & $-0.89$      & $-0.8$      \\
      &                      & Best double \gls{EMS}  & $-0.90$  & $-0.84$      & $-0.93$      & $-0.88$      \\ 
    \midrule
    \multirow{10}{*}{\shortstack{Positioning\\Error (m)}}
      & \multirow{5}{*}{AE}
        & No \gls{EMS}           & $24.5$   & $52.5$   & $17.1$      & $51.2$      \\
      &                      & Specular \gls{EMS}s    & $13.5$   & $39.0$   & $12.8$      & $33.7$      \\
      &                      & \gls{EMS}1 only (best) & $13.7$   & $40.5$   & $13.2$      & $39.7$      \\
      &                      & \gls{EMS}2 only (best) & $16.3$   & $45.7$   & $10.6$      & $39.0$      \\
      &                      & Best double \gls{EMS}  & $11.0$   & $22.5$   & $9.3$      & $18.0$      \\ 
    \cmidrule(l){2-7}
      & \multirow{5}{*}{\gls{tsne}}
        & No \gls{EMS}           & $17.5$   & $61.5$   & $16.4$      & $58.5$      \\
      &                      & Specular \gls{EMS}s    & $13.2$   & $30.0$   & $11.9$      & $25.5$      \\
      &                      & \gls{EMS}1 only (best) & $14.7$   & $48.7$   & $13.35$      & $39.7$      \\
      &                      & \gls{EMS}2 only (best) & $14.4$   & $36.0$   & $13.8$      & $33.7$      \\
      &                      & Best double \gls{EMS}  & $11.0$   & $22.5$   & $9.8$      & $20.2$      \\ 
    \bottomrule
  \end{tabular}
\end{table*}

\subsection{Joint Impact of \gls{SNR}, Dissimilarity, and \gls{ris} Comparison}

\begin{figure}[t]
  \centering
  \includegraphics[width=.95\linewidth]{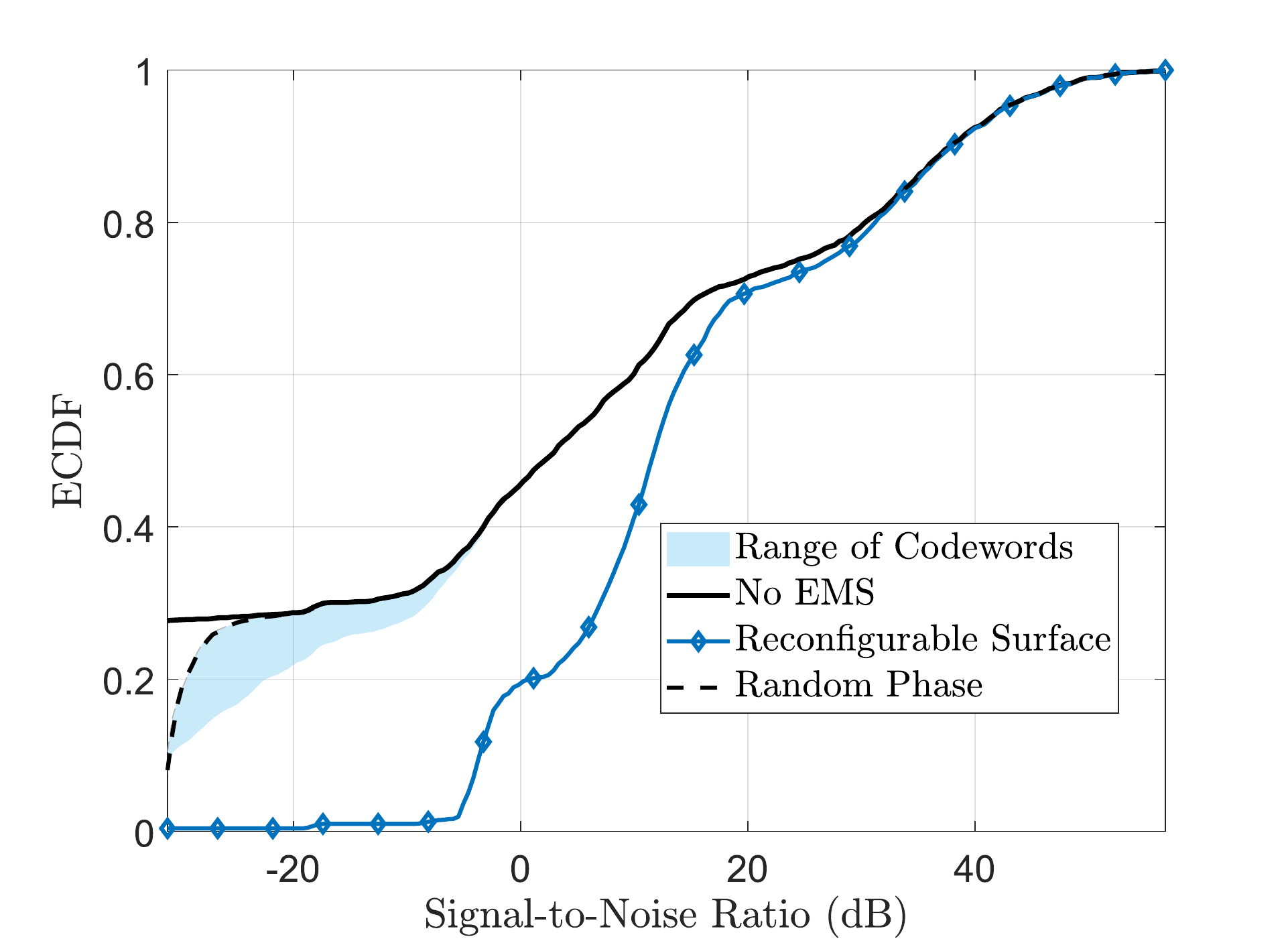}\\[1mm]
  \includegraphics[width=.95\linewidth]{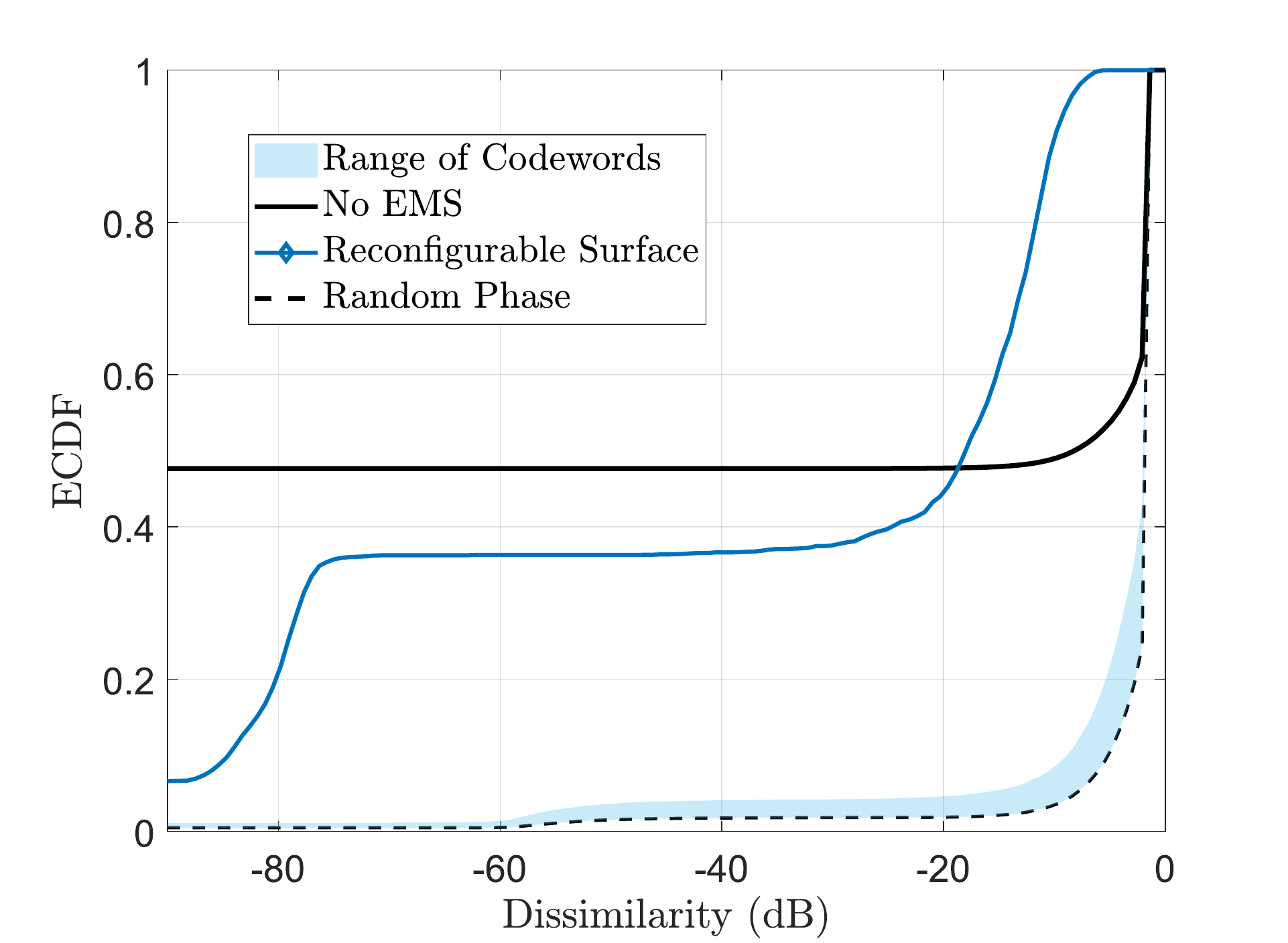}\\[1mm]
  \includegraphics[width=.95\linewidth]{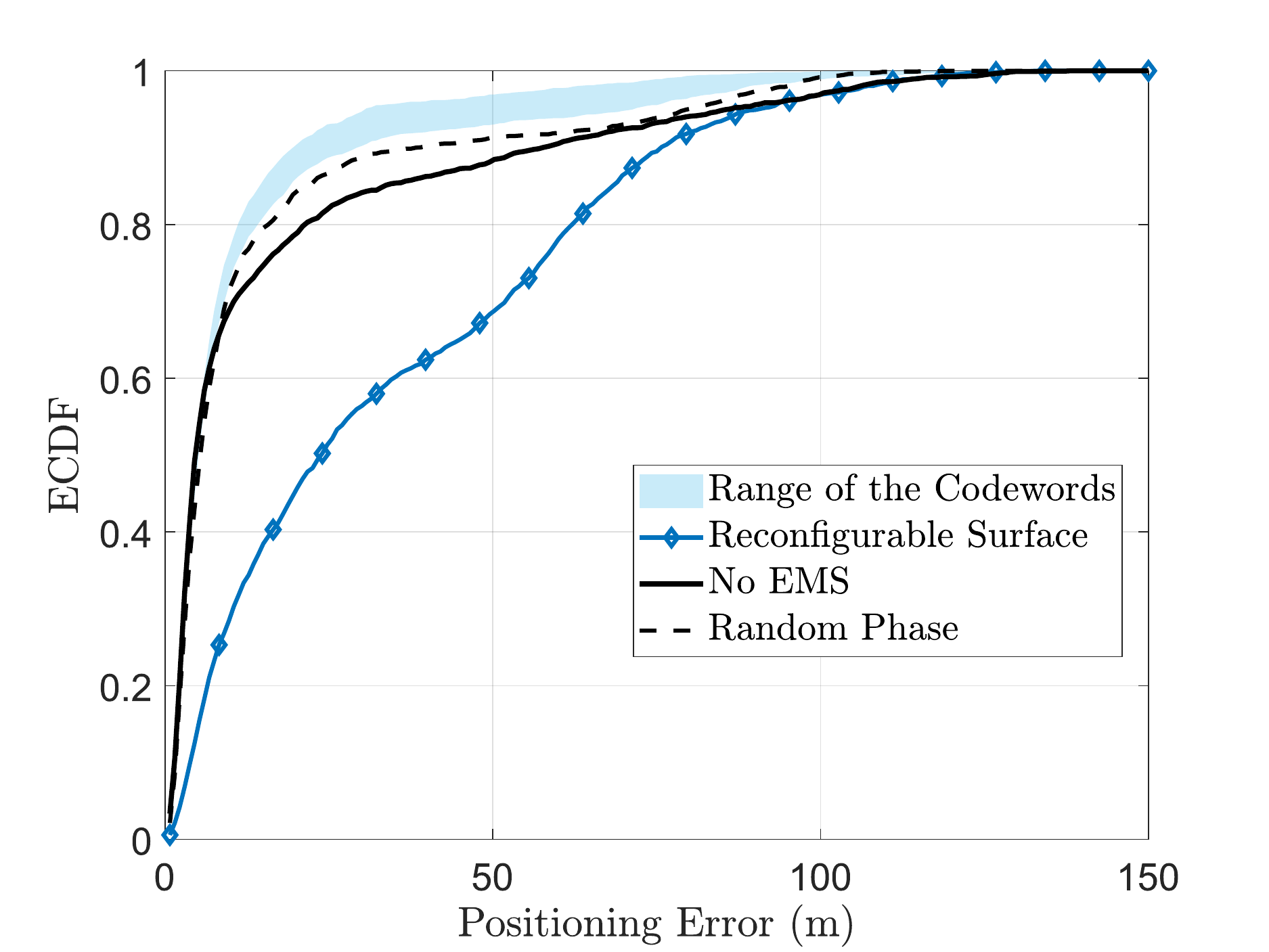}
  \caption{Empirical CDFs of \gls{SNR} (top), \gls{LE}-based dissimilarity (middle), and positioning error (bottom) for no \gls{EMS}, static codebook \gls{EMS}s, idealized \gls{ris}, and random phase panels.}
  \label{fig:ECDF_Comparison}
\end{figure}

Figure~\ref{fig:ECDF_Comparison} further explores how \gls{SNR}, \gls{LE}-based dissimilarity, and positioning error vary under four EMS deployment strategies: no \gls{EMS}, static codebook \gls{EMS}s, an idealized reconfigurable \gls{ris} (which is impractical, as it requires real-time user location knowledge), and random static phase.

The reconfigurable \gls{ris} achieves the highest \gls{SNR}, but it \textit{degrades} the embedding and error performance since it eliminates location-dependent channel diversity by always maximizing gain in a fixed direction. Random static phase provides high channel diversity, but suffers from poor \gls{SNR}, which makes localization unreliable. Codebook-optimized \gls{EMS}s, instead, balance \gls{SNR} and diversity, achieving the lowest error among all practical schemes.

\subsection{Discussion and Design Insights}

The improvements from codebook-based \gls{EMS}s are most significant for user positions in the 60th–95th percentiles of error distribution, i.e., for the hardest-to-localize users in \gls{nlos} regions. For the remaining (mostly \gls{los}) positions, gains are minor, as the system is already close to optimal. Importantly, the amount of improvement is scenario-dependent and is closely linked to both the \gls{EMS} position and codebook choice. Optimal performance would require joint network planning and \gls{EMS}/codebook co-design, which is left for future work.

From a computational perspective, the offline nature of the EMS design process allows the use of large codebooks and many panels, at the cost of higher (but not real-time) simulation complexity. The online part (charting and localization) is not affected by these design choices.

Finally, comparison with reconfigurable \gls{ris} reveals a key insight: maximizing \gls{SNR} alone does not guarantee good localization, as it suppresses the spatial fingerprints required by \gls{cc}. Instead, the introduction of carefully chosen static multipath (via \gls{EMS}s) improves both the \gls{TW} and \gls{CT} of the embedding and substantially reduces large errors.
\glsresetall

\section{Conclusion}
This paper presented a comprehensive framework for enhancing channel charting (CC) through the use of static electromagnetic surfaces (EMS) in realistic urban environments. By enriching the multipath structure of the propagation channel, the proposed EMS-assisted setup improves the geometric consistency of the channel-state features without requiring any form of active reconfiguration or prior user knowledge.  

Two complementary CC techniques—semi-supervised t-SNE and a semi-supervised autoencoder—were jointly employed to validate the reliability of the learned embeddings across nonparametric and parametric mappings. Both methods consistently demonstrated that static, codebook-optimized EMSs substantially reduce localization errors and improve trustworthiness and continuity, particularly for users in non-line-of-sight regions. 

A detailed comparison with an idealized reconfigurable intelligent surface (RIS) and with random static phases revealed a fundamental insight: CC accuracy depends not on maximizing signal-to-noise ratio (SNR) or spatial dissimilarity alone, but on achieving a balanced trade-off between them. While excessive focus on SNR suppresses location fingerprints, moderate SNR combined with diverse multipath signatures yields more distinctive and stable embeddings.  

Overall, the findings confirm that physically preconfigured, passive EMSs can serve as an effective and low-cost means to enhance CC-based localization. Future research may explore joint optimization of EMS placement and phase codebooks, as well as potential extensions to dynamic or user-aware surface designs.

\bibliographystyle{IEEEtran}
\bibliography{IEEEabrv,biblio.bib}

\end{document}